\newif\ifarxiv
\arxivfalse

\ifarxiv
  \PassOptionsToClass{nonacm}{acmart}
\fi

\PassOptionsToPackage{prologue,dvipsnames}{xcolor}
\documentclass[sigconf, anonymous=false, review=false, 10pt]{acmart} %

\usepackage{xurl}
\usepackage{booktabs}
\usepackage[ruled]{algorithm2e}
\usepackage{algpseudocode}
\usepackage{enumerate}
\usepackage[english]{babel}
\usepackage{blindtext}
\usepackage[caption=false,font=footnotesize,subrefformat=parens]{subfig}
\usepackage{amsmath}

\usepackage{amssymb}
\usepackage{xspace}
\usepackage{multirow}
\usepackage{threeparttable}
\usepackage{float}
\usepackage{flafter}
\usepackage{comment}
\usepackage{soul}
\usepackage[skip=12pt]{caption}
\usepackage{graphicx}
\usepackage{enumitem}
\usepackage{subfig}
\usepackage{diagbox,multirow,makecell,pifont}
\usepackage{tikz}
\usetikzlibrary{positioning,arrows.meta,calc}
\usepackage[utf8]{inputenc}

\def\fig{Fig.\xspace}

\def\tab{Tab.\xspace}

\def\ie{{\textit{i.e.}\xspace}} 
\def\eg{{\textit{e.g.}\xspace}}

\newcommand{\head}[1]{{\noindent \textbf{#1:}}}

\graphicspath{{./figures/}}
\graphicspath{{./pdf/}}
\DeclareGraphicsExtensions{.pdf,.jpeg,.png}

\ifodd 0
\newcommand{\rev}[1]{{\color{blue}#1}} %
\newcommand{\revap}[1]{#1} %
\newcommand{\com}[1]{\textbf{\color{red}(COMMENT: #1)}} %
\newcommand{\todo}[1]{\textbf{{\color{orange}(TODO: #1)}}}
\else
\newcommand{\rev}[1]{#1}
\newcommand{\revap}[1]{#1}
\newcommand{\com}[1]{}
\newcommand{\todo}[1]{}
\fi

\newcommand{\bs}[1]{\boldsymbol{#1}}

\newcommand{\scite}[1]{\textsuperscript{\tiny\cite{#1}}}

\usepackage{etoolbox}

\newcommand{\defineterm}[3]{%
  \newcounter{#2Counter}%
  \expandafter\newcommand\csname #2\endcsname{%
    \ifnumequal{\value{#2Counter}}{0}{%
        \stepcounter{#2Counter}%
        \ifx&#3&%
          #1\ (#2)%
        \else
          #1\ (#3)%
        \fi
        }{%
        \ifx&#3&%
          #2%
        \else
          #3%
        \fi
        }%
        \xspace%
  }%
  \expandafter\newcommand\csname f#2\endcsname{#1\xspace}%
  \expandafter\newcommand\csname a#2\endcsname{%
    \ifx&#3&%
      #2%
    \else
      #3%
    \fi
    \xspace
  }
  \expandafter\newcommand\csname fa#2\endcsname{%
    \ifx&#3&%
      #1\ (#2)%
    \else%
      #1\ (#3)%
    \fi%
    \xspace
  }
  \gappto{\sectionreset}{\setcounter{#2Counter}{0}}%
}

\newcommand{\sectionreset}{}

\let\oldsection\section
\renewcommand{\section}{\sectionreset\oldsection}

\defineterm{Wilderness Search and Rescue}{WiSAR}{}

\defineterm{Luneburg Lens}{LL}{}
\defineterm{Polylactic Acid}{PLA}{}
\defineterm{Fused Deposition Modeling}{FDM}{}
\defineterm{stereolithography}{SLA}{}
\defineterm{Flexible Printed Circuits}{FPC}{}
\defineterm{Auto Gain Control}{AGC}{}
\defineterm{Received Signal Strength}{RSS}{}
\defineterm{Received Signal Strength indicator}{RSSI}{}
\defineterm{Network Interface Card}{NIC}{}
\defineterm{Access Point}{AP}{}
\defineterm{Service Set Identifier}{SSID}{}
\defineterm{Pre-Shared Keys}{PSK}{}

\defineterm{Known Beacon Attack}{KBA}{}
\defineterm{Traffic Indication Map}{TIM}{}
\defineterm{Raspberry Pi Compute Module 4}{RPICM}{RPI-CM4}
\defineterm{Global Navigation Satellite System}{GNSS}{}

\defineterm{Ultra Wide Band}{UWB}{}
\defineterm{Angle of Arrival}{AoA}{}

\defineterm{electromagnetic}{EM}{}

\defineterm{Fast Fourier Transform}{FFT}{}
\defineterm{Multiple Signal Classification}{MUSIC}{}
\defineterm{Bidirectional Cyclic Coordinate}{BCC}{}

\usepackage{xifthen} %

\ifarxiv
  \setcopyright{none}
  \renewcommand\footnotetextcopyrightpermission[1]{}
\else
\makeatletter
\def\ACM@cc@type{by}
\makeatother

\acmYear{2026}\copyrightyear{2026}
\setcopyright{cc}
\setcctype[4.0]{by}
\acmConference[MobiCom '26]{The 32nd Annual International Conference on Mobile Computing and Networking}{October 26--30, 2026}{Austin, TX, USA}
\acmBooktitle{The 32nd Annual International Conference on Mobile Computing and Networking (MobiCom '26), October 26--30, 2026, Austin, TX, USA}
\acmDOI{10.1145/3795866.3796679}
\acmISBN{979-8-4007-2505-0/26/10}

\fi

\newcommand{\sysname}{\textsf{\textsc{Wi\textsuperscript{2}SAR}}\xspace}

\begin{document}

\ifarxiv
\title[\sysname]{"Take Me Home, Wi-Fi Drone": A Drone-based Wireless System for Wilderness Search and Rescue%
\texorpdfstring{\thanks{This is an author-prepared preprint. The final version will appear in the ACM MobiCom'26 proceedings.}}{}}
\else
\title[\sysname]{"Take Me Home, Wi-Fi Drone": A Drone-based Wireless System for Wilderness Search and Rescue}
\fi

\author{Weiying Hou}
\affiliation{ 
    \institution{The University of Hong Kong}
    \city{}\country{}
}
\email{weiying.hou@connect.hku.hk}

\author{Luca Jiang-Tao Yu}
\affiliation{ 
    \institution{The University of Hong Kong}
    \city{}\country{}
}
\email{lucayu@connect.hku.hk}

\author{Chenshu Wu}
\affiliation{ 
    \institution{The University of Hong Kong}
    \city{}\country{}
}
\email{chenshu@cs.hku.hk}

\renewcommand{\shortauthors}{Hou et al.}
\renewcommand{\authors}{Weiying Hou, Luca Jiang-Tao Yu, Chenshu Wu}

\newcommand{\blueurl}[1]{
\href{#1}{\textcolor{blue}{\texttt{#1}}}\xspace
}
\def\codeurl{\blueurl{https://github.com/aiot-lab/Wi2SAR}}
\def\weburl{\blueurl{https://aiot-lab.github.io/Wi2SAR}}

\begin{abstract}

Wilderness Search and Rescue (WiSAR) represents a longstanding and critical societal challenge, demanding innovative and automatic technological solutions. 
In this paper, we introduce \sysname, a novel autonomous drone-based wireless system for long-range, through-occlusion WiSAR operations, without relying on existing infrastructure. 
Our basic insight is to leverage the automatic reconnection behavior of modern Wi-Fi devices to known networks.
By mimicking these networks via on-drone Wi-Fi, \sysname uniquely facilitates the discovery and localization of victims through their accompanying mobile devices. 
Translating this simple idea into a practical system poses substantial technical challenges. 
\sysname overcomes these challenges via three distinct innovations: 
(1) a rapid and energy-efficient device discovery mechanism to discover and identify the target victim, 
(2) a novel RSS-only, long-range direction finding approach using a 3D-printed Luneburg Lens, amplifying the directional signal strength differences and significantly extending the operational range, 
and 
(3) an adaptive drone navigation scheme that guides the drone toward the target efficiently. 
We implement an end-to-end prototype and evaluate \sysname across various mobile devices and real-world wilderness scenarios. Experimental results demonstrate \sysname's high performance, efficiency, and practicality, highlighting its potential to advance autonomous WiSAR solutions. 
\sysname is open-sourced at \weburl to facilitate further research and real-world deployment.

\end{abstract}



\maketitle

\section{Introduction}
\label{sec:introduction}

Wilderness Search and Rescue (WiSAR) has become an increasingly critical societal challenge due to the rising popularity of outdoor activities, particularly mountaineering~\cite{hansenOutdoorRecreationSweden2023,derksCOVID19inducedVisitorBoom2020}. 
Missing person incidents have increased markedly, \eg, cases in England and Wales have grown by an average of 11\% annually since 2016~\cite{whitesideMountainRescueWINTER2024}.
Victims are often unable to rescue themselves due to the high injury rate in these incidents~\cite{whitesideMountainRescueWINTER2024}. Furthermore, despite carrying mobile phones, they are frequently incapable of calling for help because of weak or unavailable cellular, GPS, and satellite connectivity in remote, forested terrain~\cite{daceyUnderstandingLostPerson2023,moorePredictingGPSFidelity2023,albaneseSARDOAutomatedSearchandRescue2022}.
Consequently, many missing person cases are reported by their families or friends only after the individual is overdue.
These circumstances underscore the critical importance of timely rescue, compelling rescue teams to act swiftly to precisely locate the missing person.

Traditional WiSAR methods primarily rely on ground teams conducting grid searches based on a \emph{Last Known Position} (LKP) provided by the victim's friends or family, which is time-consuming, and often inadequate in vast and complex terrain. Recently, \revap{fueled by the booming low-altitude economy (LAE) technologies,} drones have emerged as a powerful ally for WiSAR~\cite{dji_enterprise_drone_rescue}, rapidly scanning vast landscapes and dense vegetation where human-led searches
often struggle~\cite{lyuUnmannedAerialVehicles2023}.
However, existing drone-aided WiSAR systems primarily rely on RGB and thermal cameras~\cite{schedlAutonomousDroneSearch2021,tusnioEfficiencyDronesUsage2021,murphyImprovingDroneImagery2023,lyuUnmannedAerialVehicles2023,albaneseSARDOAutomatedSearchandRescue2022},
which fail under blockages from forest canopies or rugged rocky areas, precisely where disappearances are most common.
Radio frequency (RF) signals have emerged as a promising alternative as they can penetrate visual obstructions.
Prior work has explored \revap{millimeter-wave (mmWave) radar} for vital sign detection~\cite{zhangRFSearchSearchingUnconscious2023}, but such faint signals vanish in outdoor conditions, underscoring a need for new RF-based solutions for WiSAR conditions.

\begin{figure}[t]
    \centering
    \includegraphics[width=\linewidth]{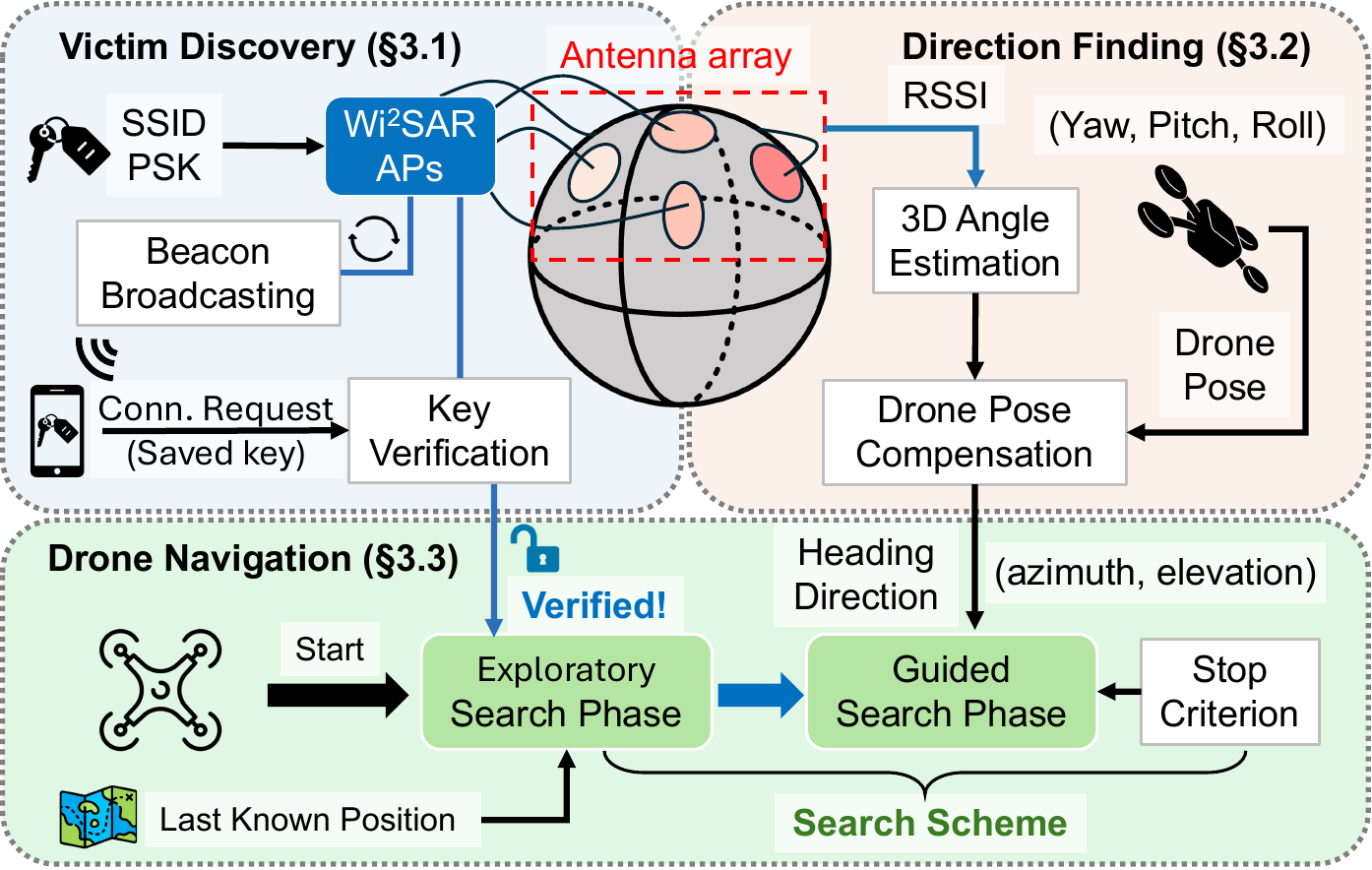}
    \caption{Overview of \sysname System.} 
    \label{fig:system_overview}
\end{figure}

In this paper, we explore and leverage a new opportunity to shift the paradigm of drone-aided WiSAR from searching human figures to locating accompanied radio devices. 
We observe that most individuals lost in the wilderness carry a mobile device (\eg, a smartphone, smartwatch, or AirTag)~\cite{daceyUnderstandingLostPerson2023}, whose signals can serve as a persistent \emph{life pulse}.
By leveraging these signals, victims can be rapidly pinpointed even beneath dense forest canopies or in rugged terrain, overcoming the limitations of conventional optical and thermal approaches and largely broadening the odds of survival through timely rescue.

Building upon this simple but effective insight, we introduce \sysname, a novel drone-based wireless system that exploits ubiquitous Wi-Fi signals for accurate and automatic victim search in wilderness environments. 
\sysname harnesses the automatic reconnection behavior of modern Wi-Fi devices, which reconnect to known networks upon detecting their beacons. 
By simulating a known Wi-Fi Access Point (\eg, home router) on the drone, \sysname can elicit identifiable packets from the target devices, effectively serving as \emph{life signals} that navigate the drone to progressively converge on the victim's location. 
Despite this straightforward principle, translating it into a practical and deployable system poses substantial challenges:

\head{$\bullet$ Victim Discovery}
Victim discovery in a relatively large area without an exact location is the foremost challenge in WiSAR, where the primary objective is to detect the victim's device as quickly and from as great a distance as possible. 
Achieving this with standard Wi-Fi networks in the context of WiSAR is non-trivial, since Wi-Fi radio on commodity smartphones typically communicates with a range of tens of meters, adequate for most indoor applications but not for WiSAR scenarios. As it is impossible to apply any changes to the victim's Wi-Fi devices in advance, how to passively increase the communication range in a \emph{non-cooperative} way calls for an innovative design. 

\head{$\bullet$ Target Localization} 
Building on the discovery phase, once the victim's device is locked onto, the drone must rapidly approach the target to narrow the search space and maintain reliable connectivity, demanding a robust localization solution in a standard Wi-Fi network. Traditional Wi-Fi trilateration techniques based on multiple range measurements are inadequate for this scenario, as they typically rely on external anchor infrastructure and can hardly guide the drone towards the target.
Therefore, accurate direction finding in 3D space is critical to ensure that the drone continually closes in on the victim and stays within the communication range, rather than flying \emph{away} from it. 
Despite extensive research on \revap{Angle-of-Arrival (AoA)} estimation using Wi-Fi signals, 
they generally focus on short-range, controlled 2D indoor environments with carefully arranged and calibrated phased arrays and known anchors as references~\cite{kotaruSpotFiDecimeterLevel2015,xiongArrayTrackFineGrainedIndoor2013,gjengsetPhaserEnablingPhased2014,pizarroAccurateUbiquitousLocalization2021}.
Unfortunately, these assumptions do not hold for WiSAR, where the drone, flying high above the ground, must perform 3D direction finding over very long distances with signals barely above the noise floor using an in-motion antenna array.

\head{$\bullet$ Drone Navigation} 
After establishing the victim's direction, the drone must translate these angular estimates into navigation commands that progressively refine its flight trajectory. Ensuring uninterrupted communication coverage is essential to avoid disconnection and unnecessary delays. Moreover, efficient navigation reduces the number of packets that need to be exchanged with the victim's device, indirectly helping to conserve its limited battery, which is critical in life-or-death situations. Implementing \sysname as a real-time, end-to-end system on a commodity drone introduces additional layers of complexity.

As outlined in \fig\ref{fig:system_overview}, \sysname addresses these challenges through three novel modules.

\head{$\blacksquare$ Long-Range Victim Discovery (\S\ref{ssec:dsn_victim_discovery})}
To enable reliable victim discovery at long distances, \sysname leverages a 3D-printed \fLL, a gradient-index spherical lens that concentrates incident signals on its surface and amplifies both downlink and uplink transmissions. \sysname periodically broadcasts beacon frames for a known network. This network is defined by its Service Set Identifier (SSID), the name of the Wi-Fi network, and credentials, such as a Pre-Shared Key (PSK), both of which are provided when the incident is reported, enabling the system to trigger an authenticated auto-reconnection from the victim's device and elicit identifiable packets. By combining amplification and authentication, \sysname can reliably identify the victim's signal from background noise, even at significant distances.

\head{$\blacksquare$ Amplitude-Only 3D AoA Estimation (\S\ref{ssec:dsn_direction_finding})}
Beyond range amplification, \sysname exploits the \fLL's focusing property to enable direction estimation without a calibrated phased array. Concentrated signals on distinct surface regions encode spatial angles as measurable \RSS patterns, allowing a lightweight RSS-only algorithm to extract both azimuth and elevation from a single packet. By eliminating the need for phase calibration, the design remains robust on a flying platform and effective even where conventional Wi-Fi arrays fail.

\head{$\blacksquare$ Direction-Guided Drone Navigation (\S\ref{sec:search-scheme})}
\sysname executes a dual-phase search scheme. It begins with an exploratory grid-based phase to cover a large area and discover the target. Once the victim's device signal is detected and verified, the system transitions into a direction-guided approach phase, where the drone continuously refines its heading using reliable direction estimates. The search terminates upon meeting a stop criterion determined by the estimated 3D direction, indicating the target is directly below, where the drone reports the final location to the ground rescue team.

\head{Summary of results} We prototype \sysname on a commercial drone platform by integrating a low-cost 3D-printed Luneburg Lens array with commodity Wi-Fi modules. The system runs in real time, feeding live AoA estimates into drone navigation to achieve an end-to-end WiSAR pipeline. We evaluate it in four representative environments with varying forest coverage and terrain complexity. Results show up to 104\% extension of the working range compared to a traditional antenna array without a Luneburg Lens on the 5\,GHz band, robust 3D direction estimation with a median angular error of 18.4$^\circ$, and efficient search over 160,000 m$^2$ within \rev{13.5} minutes with 100\% discovery rate of target devices. In a field-like WiSAR case study in a forested area of 40,000 m$^2$, \sysname discovers the victim's device within 4 minutes and reports a final localization error of 5 m. These outcomes highlight \sysname's strong potential for practical deployment.

\head{Contribution} Our core contributions are as follows:
\begin{itemize}[leftmargin=*]
    \item We design \sysname, the first automatic WiSAR system using the drone-based Wi-Fi network to discover and locate a victim without any infrastructure support.
    \item We propose an RSS-only, long-range 3D AoA estimation method built on a 3D-printed Luneburg Lens, significantly extending the operational range.
    \item We implement \sysname as an integrated real-time system on a consumer drone and validate its performance through extensive experiments in realistic mountain environments.
\end{itemize}

\begin{figure}[t]
    \centering
    \includegraphics[width=\linewidth]{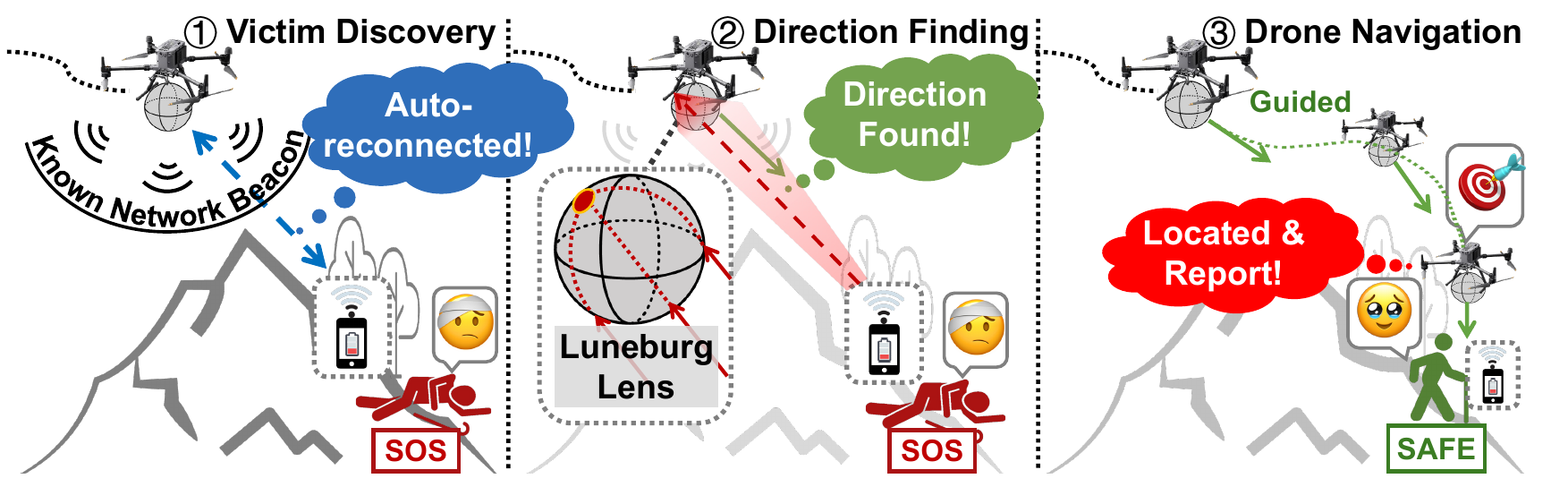}
    \caption{Typical Working Scenario.}
    \label{fig:overview}
\end{figure}

\section{\sysname Scope and Design Choices}
\label{sec:scope}
WiSAR missions can vary significantly across cases. In \sysname, we focus on a representative subset of common scenarios where a missing person is reported by an emergency contact  (\eg, a friend or relative of the victim) usually because the individual is overdue~\cite{whitesideMountainRescueWINTER2024,daceyUnderstandingLostPerson2023}. 
Rather than attempting to address all possible situations, \sysname is designed as a best-effort solution aimed at maximizing applicability across diverse cases to save lives. 

\head{Scope} Again, we are motivated by the goal of maximizing the chance of survival through timely rescue, rather than covering all WiSAR scenarios, which is practically impossible. Our current design leverages the auto-reconnection behavior of Wi-Fi under several conditions: 

\revap{\noindent\textit{1) LKP and known network information:} In typical scenarios involving an emergency report, the emergency contact can provide the victim's \emph{Last Known Position} (LKP), \eg, the last message/picture indicating the victim's hiking destination, as well as the credentials\footnote{\revap{If unavailable, \sysname falls back to utilizing public Wi-Fi (see \S\ref{sec:discuss}).}} of a known Wi-Fi network (\ie, SSID and PSK of the victim's home router), which are typically stored in plain text on mainstream mobile devices of the reporter. In practice, this process can be streamlined via an official emergency App;}

\noindent\textit{2) Powered devices:} We assume that the victim's smartphone or other Wi-Fi gadgets remain operational with Wi-Fi enabled. While this assumption does not apply to all the cases, it covers a significant portion of them, \eg, when the victim gets injured and/or lost and cannot return, his/her phone will still be functional given that modern phones have long battery life and a majority of users keep Wi-Fi constantly on for convenience\footnote{Our survey of 115 users on phone usage behaviors shows that around 78.1\% users keep Wi-Fi constantly activated during outdoor activities and the ratio goes to 91.0\% in daily life.};

\noindent\textit{3) Device proximity:} The victim's smartphone will not be far from the victim, if not co-located with them;

\noindent\textit{4) WiSAR scope:} The current \sysname mainly targets victim search (\ie, discovery and localization), and the rescue operations after finding the victim are out of our scope;

\noindent\textit{5) Single drone:} We primarily focus on single-victim search using a single drone. 
Nevertheless, \sysname naturally extends to multiple victims, \eg, through SSID-multiplexed identification and dynamic prioritization of discovered devices;
\sysname also extends to multiple drones, yet we keep it as future work to explore optimized multi-drone coordination.

\head{Typical Working Scenario}
As illustrated in \fig\ref{fig:overview}, \sysname integrates its modules in a typical wilderness search case. Consider a hiker who gets lost in a dense forest while mountaineering. The emergency contact reports the incident to the local search and rescue team, providing both the victim's \emph{Last Known Position} (LKP) and the SSID and password of the victim's home Wi-Fi network. With this information, \sysname configures a drone to mimic the known AP and initiates a grid-based exploratory flight centered around the LKP. During this phase, the \textit{Victim Discovery Module} (\S\ref{ssec:dsn_victim_discovery}) continuously broadcasts beacon frames of the specified SSID. When the victim's smartphone detects the beacons and attempts to reconnect, identifiable traffic is generated. The \textit{Direction Finding Module} (\S\ref{ssec:dsn_direction_finding}) then estimates both azimuth and elevation of the signals, even under occlusions such as thick foliage, enabling the \textit{Drone Navigation Module} (\S\ref{sec:search-scheme}) to adjust the drone's heading toward the victim. Once the drone is nearly overhead, indicated by an elevation angle approaching 90$^\circ$, \sysname pinpoints the victim's location.

\begin{table}[t]
\centering
\scriptsize
\caption{Comparison of RF Signals Against WiSAR Req.}
\begin{tabular}{l l l l}
\toprule
\textbf{Signal} & \textbf{Range} & \textbf{ID Type} & \textbf{Limitations \& Req. violated} \\
\midrule
SatCom     & Regional & Caller ID & Limited device/region; need LoS\scite{maLEOSatelliteNetwork2024} \ding{182}\ding{186} \\
GNSS       & Global & (RX only) & Need active sharing, and LoS\scite{moorePredictingGPSFidelity2023a} \ding{183}\ding{186} \\
UWB        & 5--10 m & Device ID & Very short range; need LoS\scite{chenExploitingAnchorLinks2024} \ding{186}\\
Cellular   & km-level & IMEI/IMSI & Need carrier/police coordination\scite{albaneseSARDOAutomatedSearchandRescue2022} \ding{184}\ding{186}\\
BLE        & 10--100 m & MAC addr. & Randomized MAC\scite{zehnerPrivacyEnhancingTechnologiesPhysicalLayer2025} \ding{185}\ding{186} \\
Wi-Fi      & $\sim$100 m & MAC addr. & Randomized MAC\scite{fenskeThreeYearsLater2021} \ding{185}\ding{186} \\
\bottomrule
\label{tab:signal_cmp}
\end{tabular}
\end{table}

\head{Why Wi-Fi Signals?}
Modern mobile phones are equipped with multiple wireless technologies, including satellite communication (SatCom), GNSS, ultra-wideband (UWB), cellular networks (\eg, 4G-LTE/5G-NR, 6G), Bluetooth Low Energy (BLE), and Wi-Fi. To select which of these signals can serve as a practical beacon in drone-aided WiSAR, we outline five requirements. \ding{182} The signal should rely on native protocols so that it works on most devices without additional applications or modifications. \ding{183} It cannot assume the lost person is able to cooperate, since they may be injured or unconscious~\cite{whitesideMountainRescueWINTER2024}. \ding{184} The signal must enable direct identification without carrier or police involvement, as such procedures are slow.
\ding{185} The identifier must be persistent and uniquely bound to the victim or their device, rather than anonymized (\eg, randomized). \ding{186} The effective range should be sufficiently long to penetrate wilderness obstacles, and the signal must also provide reliable directional cues for drone navigation.
\tab\ref{tab:signal_cmp} compares wireless signals against these requirements. 

While existing signals fail to meet all requirements directly, Wi-Fi offers unique potential. Although its identifiers, MAC addresses, are usually randomized~\cite{fenskeThreeYearsLater2021}, we observe that most Wi-Fi devices automatically reconnect to previously saved networks upon detecting their beacons. This auto-reconnect behavior provides an opportunity to \emph{natively} elicit \emph{persistent} and \emph{identifiable} signal from \emph{non-cooperative} devices. Fortunately, this network information can be provided at the time of reporting by the lost person's family or friends, who share the same network with the victim, satisfying requirements \ding{182}–\ding{185}. \revap{Beyond connectivity, this credential-based authentication distinguishes the target, helping to filter out signals from other active devices, such as those of rescue personnel or bystanders.} 
However, Wi-Fi still falls short of requirement \ding{186}: Its practical range is limited under forest or rocky occlusions, and accurate AoA estimation is difficult on an in-motion drone using commodity NICs. These challenges motivate \sysname's design in \S\ref{sec:design}, where we introduce three modules that jointly address both the long-range limitation and the AoA challenge for drone-based WiSAR.

\section{\sysname Design}
\label{sec:design}

\subsection{Victim Discovery}
\label{ssec:dsn_victim_discovery}
The vast search area is one of the most critical challenges in WiSAR. Search and rescue teams can often obtain the victim's \emph{Last Known Position} from emergency contacts, but this only provides a vague starting point. In practice, the search area may span hundreds of square miles~\cite{murphyImprovingDroneImagery2023}. The terrain is often rugged and covered with dense forests, which further compounds the challenge and makes it nearly impossible for manpower alone to cover every corner. More critically, the so-called \emph{golden hour} in WiSAR, when the victim's survival chances are highest, demands rapid discovery. Confirming only the presence of a victim in a certain region can dramatically narrow the search scope and accelerate rescue operations. Thus, detecting a victim swiftly within a broad area, even without precise coordinates, is often more urgent than immediately pinpointing the exact location.

We observe a promising opportunity for rapid victim discovery through Wi-Fi enabled devices that almost always accompany individuals~\cite{daceyUnderstandingLostPerson2023}.
If these devices can be reliably authenticated, \sysname can confirm a victim's presence quickly even at long range. At first glance, periodic probe requests generated by smartphones during network scanning may appear to be useful evidence of nearby devices~\cite{Freudiger2015HowTI}. However, it is infeasible to obtain the hardware-specific MAC address of the victim's accompanying device; more importantly, recent large-scale studies~\cite{fenskeThreeYearsLater2021,bravenecUJIProbesDataset2023} show that modern smartphones frequently randomize their MAC addresses during probing, which makes MAC-based identification unreliable.

\head{Auto-reconnection Behavior} 
Our approach instead leverages the auto-reconnection behavior
common to commercial Wi-Fi devices. When a device detects a known network, it automatically attempts to reconnect using stored credentials. In most home and office deployments, these credentials are based on pre-shared keys, most commonly WPA2-PSK\footnote{Although WPA/WPA2-PSK has known vulnerabilities~\cite{krack2017}, it remains the most widely used in practice. 
Our field survey shows that 99.2\% of 237 networks relied on pre-shared keys, predominantly WPA/WPA2-PSK, with a small fraction (4.6\%) using WPA3-Personal (SAE).}. 
\revap{%
We leverage this feature by configuring the drone to broadcast beacon frames that mimic a trusted private network at standard 100\,ms intervals.
This strategy relies entirely on the native Wi-Fi stack, requiring no custom App installation on the victim's device. Since it utilizes standard background scanning and authentication processes, the energy overhead is comparable to typical phone usage. 
When the victim's device initiates the authentication sequence, the standard WPA2-PSK four-way handshake confirms that the device holds the correct credentials. While this does not prove the device's unique identity, it strongly indicates that the victim is present in the search region, allowing \sysname to quickly narrow the search area to the signal coverage range.}
To the best of our knowledge, our \sysname is the first WiSAR system to leverage standard Wi-Fi auto-reconnection for victim discovery in WiSAR. We validate its feasibility through a device survey, which shows that smartphones, tablets, and smartwatches from a range of manufacturers\footnote{Our survey covers 11 smartphones from Apple, OPPO, Google, Honor, MI, and Vivo, as well as an Apple tablet and smartwatch.} consistently reconnect to familiar networks once in range, thereby enabling rapid victim identification before fine-grained localization.

\begin{figure}[t]
    \centering
    \includegraphics[width=\linewidth]{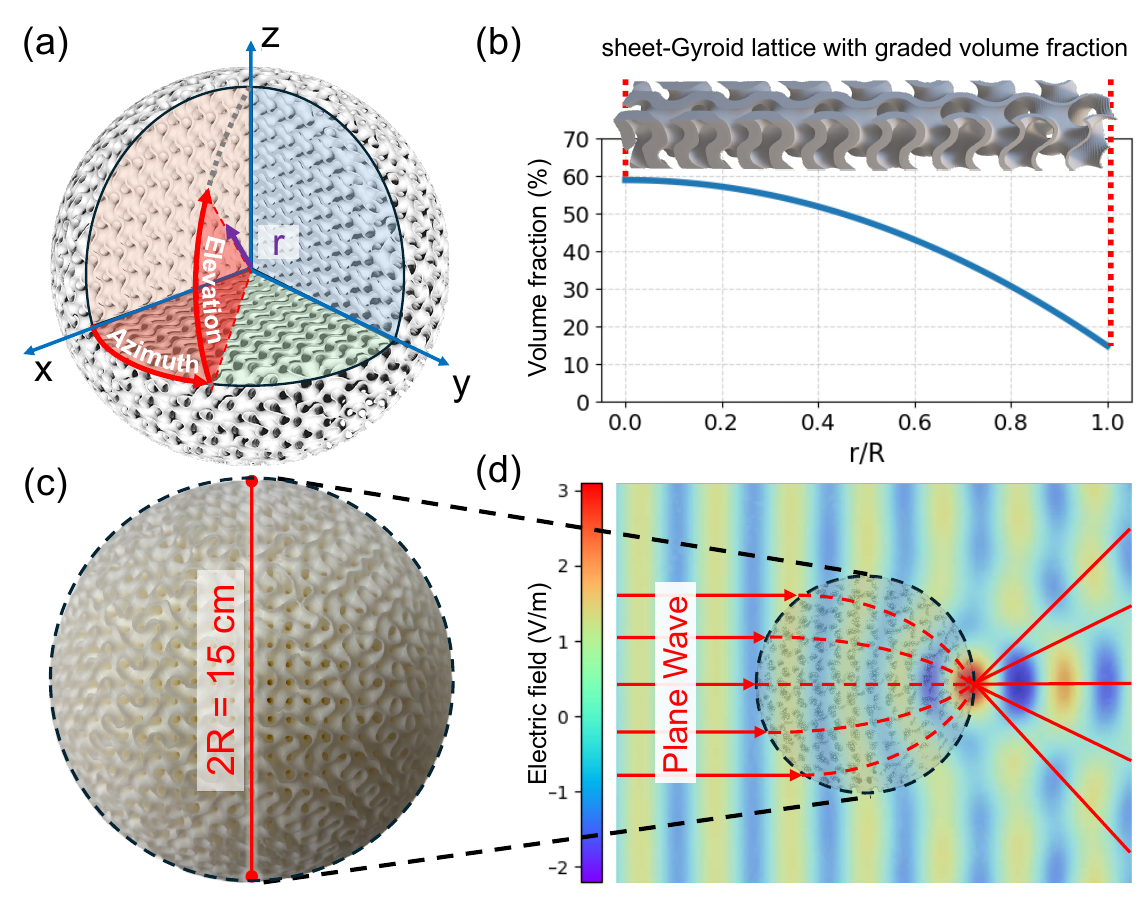}
    \caption{Design and Fabrication of the \fLL. \rm{
    (a) Cutaway showing graded sheet-Gyroid lattice with spherical coordinates. (b) Radial volume-fraction profile. (c) 3D-printed prototype, diameter 15 cm. (d) COMSOL~\cite{comsolCOMSOLMultiphysics} full-wave simulation of the 15-cm Luneburg Lens at 5.745 GHz, showing electric-field focusing at the antipode.
    }
    }
    \label{fig:ll_inner}
\end{figure}

\subsection{Direction Finding}
\label{ssec:dsn_direction_finding}
Once the target device's packet is successfully elicited and authenticated, \sysname must navigate the drone within the device's coverage robustly, and locate the device precisely and rapidly.
\revap{Conventional range-based trilateration~\cite{vasisht2016decimeter,abediNoncooperativeWifiLocalization2022a} is ill-suited for the initial search phase in WiSAR. (1) Lack of directionality: trilateration infers position from scalar ranges and does not directly provide a bearing for closed-loop navigation. Without directional guidance, the drone might inadvertently move \emph{away} from the target instead of toward it, wasting time and weakening the signal. (2) Moreover, reliable ranging is hard to obtain in WiSAR setting: time-of-flight ranging is impractical with unsynchronized transceivers~\cite{kotaruSpotFiDecimeterLevel2015, soltanaghaei2018multipath,vasisht2016decimeter}, and RSS-based path-loss models~\cite{bahlRADARBuildingRFbased2000} break down under canopy and rugged terrain. Therefore, \sysname uses AoA estimation to obtain a bearing from the outset and benchmarks against direction-based methods (\eg, AoA/triangulation) in \S\ref{ssec:exp_df}.}

Many existing AoA estimation methods rely on subspace separability such as MUSIC~\cite{xiongArrayTrackFineGrainedIndoor2013,kotaruSpotFiDecimeterLevel2015,gjengsetPhaserEnablingPhased2014}. While highly accurate indoors, they presuppose carefully calibrated phased arrays and stable 2D environments with known anchors. Our comparison study in \S\ref{ssec:exp_df} shows that an ArrayTrack-style MUSIC baseline~\cite{xiongArrayTrackFineGrainedIndoor2013} implemented on a Phaser-style array~\cite{gjengsetPhaserEnablingPhased2014} degrades by over 10$\times$ when phase drift cannot be controlled in the air. Moreover, subspace estimators fundamentally assume adequate SNR and inter-element phase coherence to keep signal and noise subspaces separable~\cite{guniaAnalysisDesignMuSiCBased2023}. In \sysname, by contrast, the first packets captured by a high-altitude drone are often close to the noise floor, making such methods unstable at long ranges. Finally, few systems perform 3D AoA, and those are typically validated only at short distances indoors~\cite{Zhang20193DWiFi3L,wangWiCALAccurateWiFiBased2025}, whereas \sysname demands robust azimuth and elevation estimation across hundreds of meters.

To overcome these challenges, we introduce a Luneburg-lens front-end and propose an amplitude-only 3D AOA estimator that works reliably at near-noise-floor SNR, avoids fragile phase synchronization, and scales well to long-range, non-cooperative WiSAR missions under aerial mobility.

\head{Luneburg Lens}
\begin{figure}[t]
    \centering
    \includegraphics[width=\linewidth]{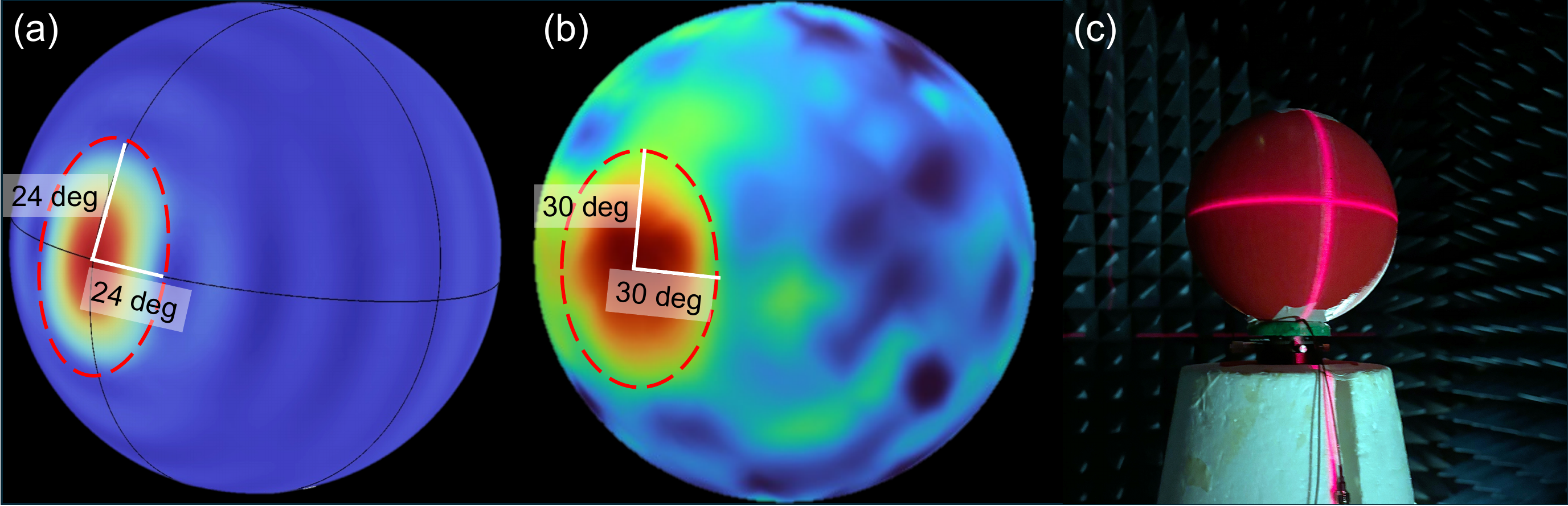}
    \caption{Luneburg Lens Beam Patterns. {
    \rm (a) COMSOL simulation, (b) measured beam, and (c) chamber setup; in (a,b), the blue–red color scale indicates signal strength.
    }
    }
    \label{fig:measure_beam_pattern}
\end{figure}
\fLL~\cite{Luneburg1966MathematicalTO} originates in optical physics and has been adapted to electromagnetics for high-gain antennas~\cite{wangHighfrequencyLimits3DPrinted2023} and mmWave backscatter retroreflectors~\cite{qianUniScatterMetamaterialBackscatter2023}. Its gradient refractive index concentrates an incident plane wave to a focal point on the opposite surface of the sphere (see \fig\ref{fig:ll_inner}d). For an ideal lens with radius $R$, the relative permittivity $\varepsilon(r)$ at radial distance $r$ follows $\varepsilon(r)=2-(r/R)^2$; equivalently, the refractive index is $n(r)=\sqrt{2-(r/R)^2}$. A practical implementation is feasible with consumer-grade 3D printers (see \fig\ref{fig:ll_inner}a–c and \S\ref{ssec:impl_luneburg_lens}).

For WiSAR, two properties of \fLL are especially valuable. First, its passive focusing provides high aperture gain with theoretical full-sphere coverage;
practically we search a hemisphere because the lens is positioned above the target. We therefore use a \emph{reference beam template} $B_0$ on the lens surface, obtained once at design time (\eg, a single far-field characterization in a microwave anechoic chamber, rather than per-deployment or on-the-fly calibration). Second, the focusing property produces a deterministic three-dimensional distribution of the electric field on the lens surface, so a plane wave $\bs{u}$ from $(\theta,\phi)$ maps to a characteristic magnitude pattern across receivers affixed to the surface. By spherical symmetry, the \emph{received beam pattern} $B_{\mathbf u}$ for an arbitrary incident direction $\bs{u}$ is a rotated version of $B_0$. We exploit this property to enable amplitude-only inference of $\bs{u}$ from RSS samples across receivers, replacing complex inter-element phase calibration with a one-time characterization.

\head{RSS-Only 3D AOA Algorithm}
Building on the above properties of the Luneburg Lens, we describe our amplitude-only 3D direction estimation method.
Since the Luneburg Lens is positioned mainly above the target, we restrict the search domain to the upper hemisphere
\begin{equation}
\Omega=\{\,(\theta,\phi)\mid \theta\in[0,\tfrac{\pi}{2}],\;\phi\in(-\pi,\pi]\,\}.
\end{equation}
Let $\bs{u}(\theta,\phi)=[\cos\theta\cos\phi,\ \cos\theta\sin\phi,\ \sin\theta]^\top$ denote the unit incident direction vector. Denote $R_{\bs{u}}$ as the rotation matrix that aligns the reference (boresight) focal direction with the incident direction $\bs{u}$. 
Let the \emph{reference beam template} on the lens surface be the continuous function $B_0:\Omega\to\mathbb{R}$ measured once in far-field conditions (see \fig\ref{fig:measure_beam_pattern}b). When a plane wave impinges from $\bs{u}$, the induced surface distribution, \ie, the \emph{received beam pattern}, is a rotated version of $B_0$
\begin{equation}
B_{\bs{u}}(\xi)=B_0\!\big(R_{\bs{u}}^{-1}\,\xi\big),\qquad \xi\in\Omega.
\end{equation}
We measure the \emph{received beam pattern} using $N$ antennas placed at known surface locations $\mathcal L=\{L_i\}_{i=1}^N\subset\Omega$.
For the $i$-th antenna, the received power satisfies \emph{Friis law}
\begin{equation}
P_{r,i}=P_t\,G_t\,G_r(\bs{u};L_i)\left(\frac{\lambda}{4\pi d}\right)^2,
\end{equation}
where $P_t$ is the transmit power (from victim's target device), $G_t$ is the TX gain\footnote{We assume TX antenna is omnidirectional, thus, $G_t$ is a constant.}, $G_r(\bs{u};L_i)=B_{\bs{u}}\,(L_i)$ is the RX directionality at location $L_i$ (considering Lens directional gain and radiation pattern of RX antenna), $d$ is the range, and $\lambda$ the wavelength.
\revap{Since RX directionality is usually measured in dB, we denote by $y_i$ the received RSS at antenna $i$ located at $L_i$ in dBm, and by $B_0^{\mathrm{dB}}(\cdot)$ the RX template in dB. We obtain the per-antenna observation model}
\begin{equation}
y_i = B_0^{\mathrm{dB}}\!\big(R_{\bs{u}}^{-1}L_i\big)\;+\;\beta\;+\;n_i,\qquad n_i\sim\mathcal N(0,\sigma^2),
\end{equation}
\revap{where direction-independent terms are absorbed into the offset $\beta$ (\eg, $P_t^{\mathrm{dBm}}$, $G_t^{\mathrm{dB}}$, distance-dependent large-scale path loss, average canopy penetration loss, and fixed RX-chain constants). The residual $n_i$ captures unmodeled fluctuations (\eg, multipath and measurement noise) and is approximated as i.i.d. Gaussian in dB. Over all $N$ antennas, we stack the measurements and residuals as}
\begin{equation}
\bs{y}=[y_1,\ldots,y_N]^\top, \bs{n}=[n_1,\ldots,n_N]^\top, \bs{n}\sim\mathcal N(\bs{0},\sigma^2\bs{I}),
\end{equation}
\revap{and define the template vector}
\begin{equation}
\bs{s}(\bs{u})=\big[B_0^{\mathrm{dB}}(R_{\bs{u}}^{-1}L_1),\ldots,B_0^{\mathrm{dB}}(R_{\bs{u}}^{-1}L_N)\big]^\top.
\end{equation}
\revap{To eliminate the unknown offset $\beta$, we remove the mean of both $\bs{y}$ and $\bs{s}(\bs{u})$ by subtracting their sample means, yielding $\tilde{\bs{y}}$ and $\tilde{\bs{s}}(\bs{u})$.}
Thus, the estimated incident direction derived from \emph{Least Squares} method can be written as:
\begin{equation}
\hat{\bs{u}}=\arg\min_{\bs{u}\in\Omega}\ \big\|\tilde{\bs{y}}-\tilde{\bs{s}}(\bs{u})\big\|_2^2=\arg\max_{\bs{u}\in\Omega}\frac{\tilde{\bs{s}}(\bs{u})^\top\,\tilde{\bs{y}}}{\|\tilde{\bs{s}}(\bs{u})\|_2\,\|\tilde{\bs{y}}\|_2}.
\end{equation}
\revap{This method has two advantages. First, it matches the measured zero-mean RSS to the template beam pattern, making it insensitive to unknown transmit power and direction-independent large-scale loss, and avoiding fragile RSS path-loss fitting. Second, it is amplitude-only: it exploits the spatial RSS pattern shaped by the Luneburg Lens without phase calibration, thereby sidestepping the synchronization requirements of phase-based AoA.}

\head{Antenna Layout Considerations}
The antenna layout $\mathcal L$ critically affects how the continuous beam pattern is discretized into the sampled vector $\bs{s}(\bs{u})$, thereby determining the resolution and robustness of our estimator. To support reliable direction estimation, we adopt a layout that emphasizes the upper hemisphere, where the Luneburg Lens concentrates most energy, and provides balanced angular sampling across azimuth and elevation (see \S~\ref{ssec:impl_luneburg_lens} for details).

\begin{figure}[t] 
\centering 
    \includegraphics[width=\linewidth]{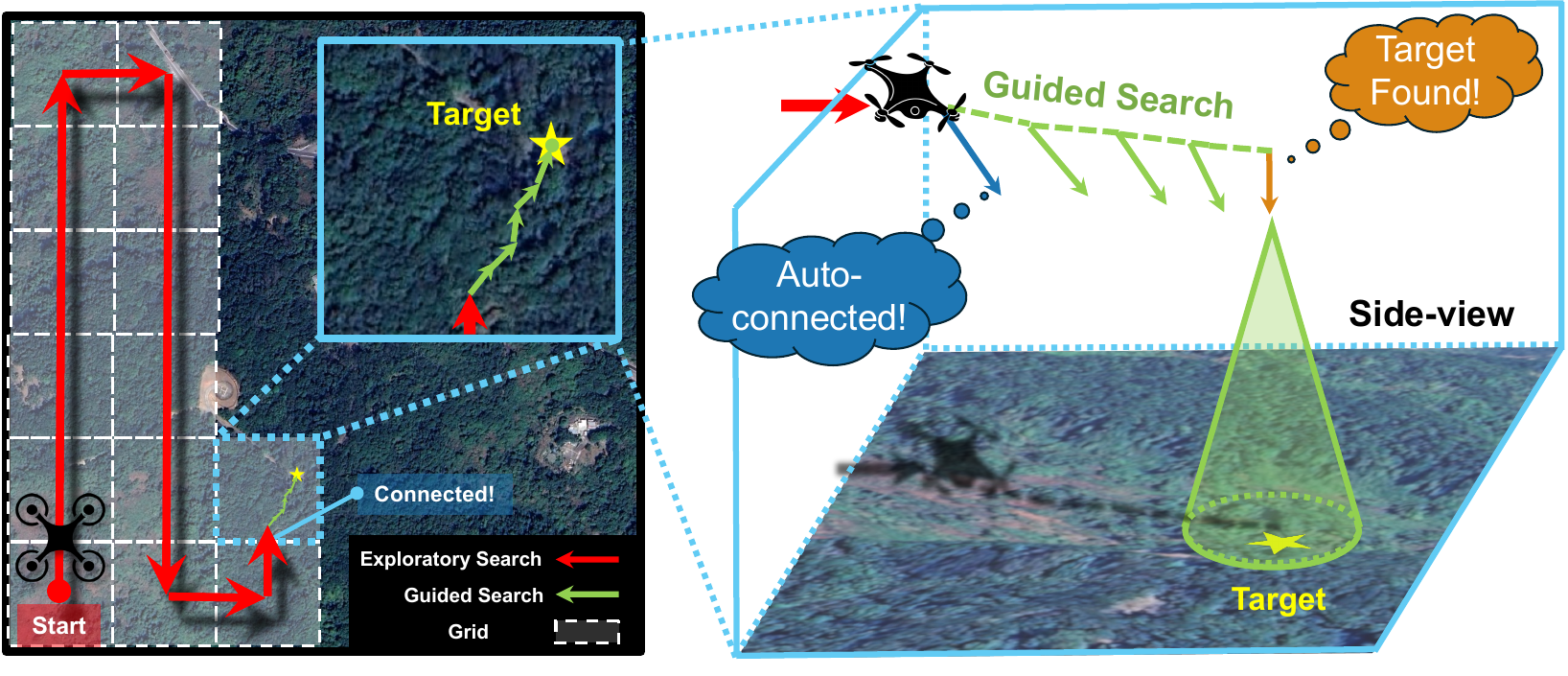} 
    \caption{Proposed Dual-Phase Search Scheme. {\rm Once the target device auto-reconnects to the on-drone AP, the search scheme shifts from~\textcolor{black}{exploratory search} to \textcolor{black}{guided search}}.}
    \label{fig:search_scheme} 
\end{figure}

\subsection{Drone Search Scheme}
\label{sec:search-scheme}
To effectively localize a victim starting from a potentially outdated or imprecise LKP provided by emergency contacts, we design our \sysname to follow a dual-phase search scheme. Our goal is to progressively resolve positional uncertainty: we first explore the entire area of interest while broadcasting known network beacons to elicit potential connection attempts, and then, once the target device is locked on, converge on the victim's device through signal-guided navigation. The process concludes once a robust geometric termination criterion is satisfied, as illustrated in \fig\ref{fig:search_scheme}.

\head{Phase 1: Exploratory Search}
The objective of this phase is to guarantee comprehensive coverage around the LKP and ensure that no potential victim device remains undetected. At the start, our drone executes a deterministic zigzag trajectory that comprehensively sweeps the uncertainty region. We design the grid spacing to be twice our system's reliable operational range, which ensures that any active device within the area will be captured by at least one flight leg. This phase ensures that the victim's signal is detected before the search narrows to fine-grained localization in the next phase.

\head{Phase 2: Guided Search}
Once our system detects and authenticates a victim's signal, \sysname transitions to the guided search phase. Here, the drone leverages our 3D direction estimates to directly navigate toward the source. This creates an iterative refinement loop: as the drone moves closer, the signal becomes stronger, which improves AoA accuracy and further accelerates convergence, leading to a more precise final localization result.

\head{Stop Criterion}
We terminate the search when the measured elevation angle surpasses a high threshold near 90$^\circ$. This criterion is robust because, close to the zenith, the elevation angle is insensitive to small horizontal displacements of the drone.
At this point, our drone records its GPS position as the victim's estimated location. While additional actions such as landing for visual confirmation are possible, they fall outside the scope of this work.

\begin{figure}[t]
    \centering
    \includegraphics[width=\linewidth]{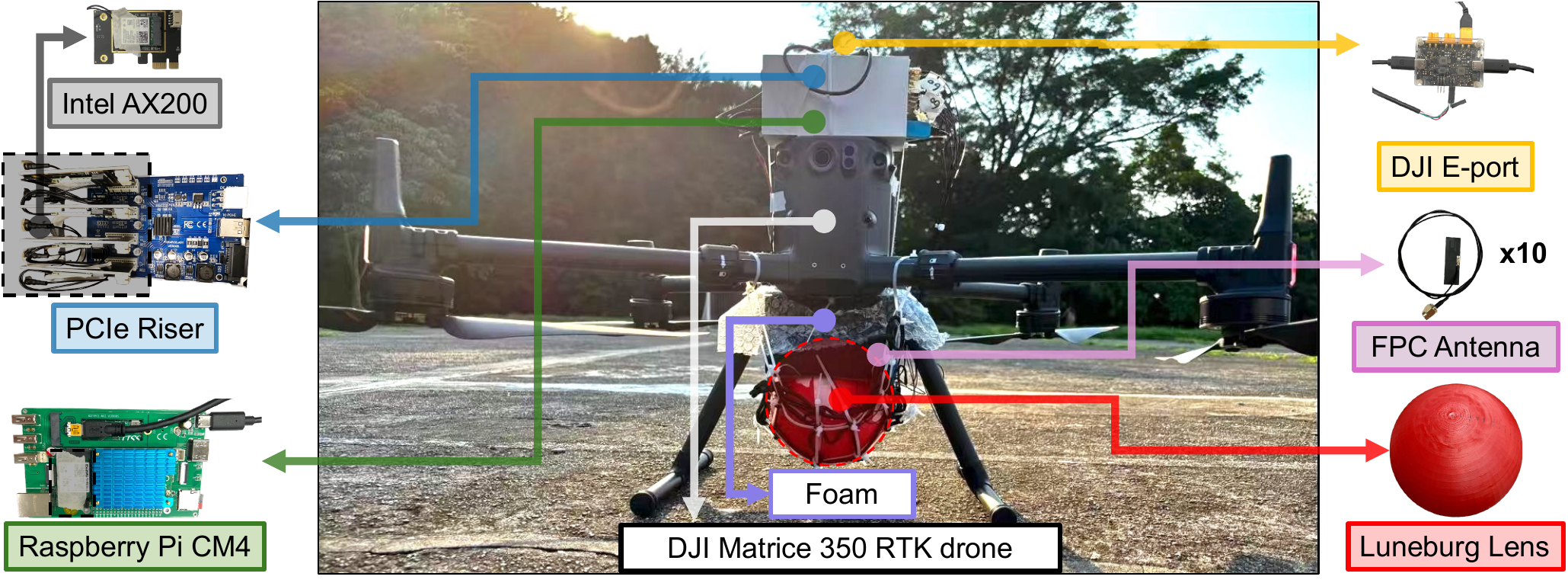}
    \caption{Prototype of \sysname.} 
    \label{fig:impl}
\end{figure}

\begin{figure}[t]
    \centering
    \includegraphics[width=\linewidth]{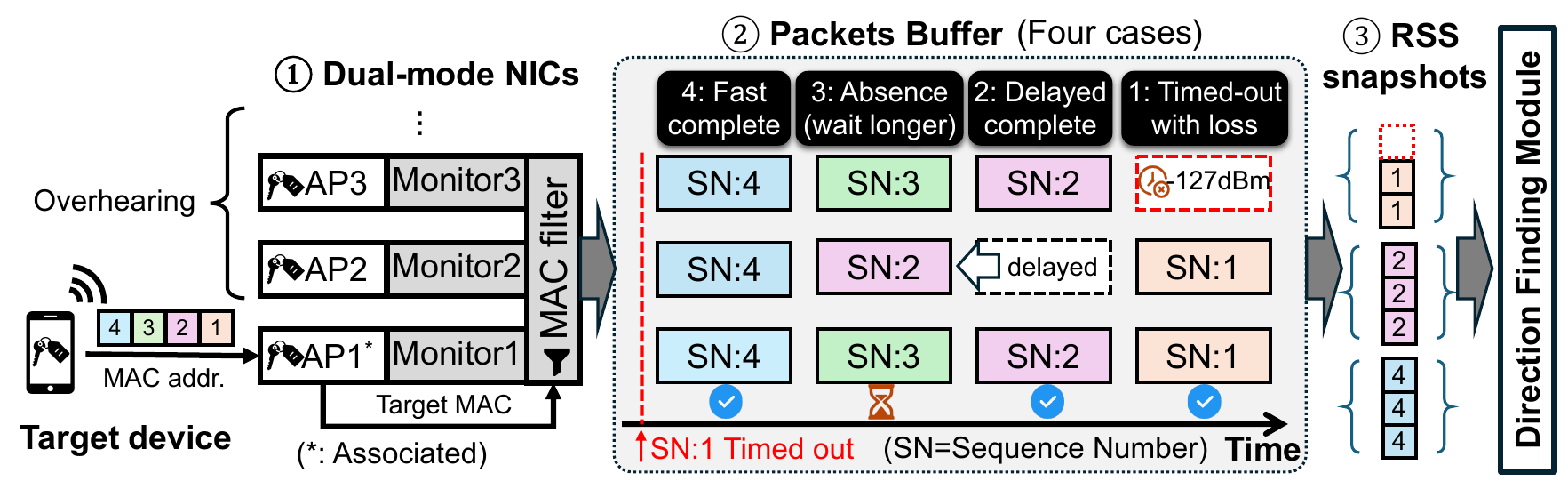}
    \caption{Victim Discovery Module. \rm{\ding{182} Dual-mode NICs lure and overhear target's uplink packets, which are MAC-filtered and buffered by sequence number. \ding{183} Packets are finalized as either complete (fast/delayed) or timed-out (with loss), while absences are transient until their timers expire. \ding{184} RSS snapshots are then extracted and forwarded to DFM.}
    }
    \label{fig:vdm_design}
\end{figure}

\section{\sysname Implementation}
\label{sec:impl}

We implement the \sysname prototype onboard a DJI Matrice~350~\cite{dji_matrice_350_rtk}, equipping the drone with a Raspberry Pi Compute Module~4~\cite{raspberry_pi_compute_module_4}. The module connects to the drone via the DJI E-port, which provides access to GPS and IMU data. As shown in \fig~\ref{fig:impl}, a 3D-printed \fLL of 15\,cm in diameter is mounted beneath the drone, while the Raspberry Pi is expanded through PCIe risers to host five Intel AX200 NICs, yielding ten antennas in total. Although multiple antennas are deployed, the system remains simple to implement because it does not require the precise phase synchronization across RF chains demanded by prior CSI- or AoA-based designs~\cite{kotaruSpotFiDecimeterLevel2015,xiongArrayTrackFineGrainedIndoor2013}. Instead, \sysname relies on packet-level RSS snapshots rather than phase coherence.
\subsection{Victim Discovery Module}
\label{ssec:impl_vdm}

The Victim Discovery Module broadcasts beacons to lure target devices for authentication while simultaneously measuring their uplink responses across the entire antenna array. This seemingly simple task is in fact non-trivial due to two fundamental challenges:
(1) \emph{Coverage alignment:} Downlink transmissions (beacons, ACKs) and uplink receptions (RSS measurements) must share not only the same angular coverage but also the same link budget. A naive design, where an omnidirectional antenna handles beaconing and the Luneburg Lens array handles reception, creates an \emph{asymmetric link}: the client may hear beacons at long range, but its responses fall outside the array's main lobe or below its decoding threshold, wasting the extended range.
(2) \emph{Coherent snapshot:} Our direction estimator requires a packet-level RSS vector, \ie, a \emph{coherent spatial snapshot} of the same MPDU across all antennas. Antenna-switching schemes that multiplex a single NIC across multiple antennas~\cite{xieSWANStitchedWiFi2018,wangWiCALAccurateWiFiBased2025} cannot provide simultaneous per-packet RSS, and are further degraded by drone motion between successive samples.

\head{Multi-NIC Dual-Mode Architecture}
Our solution (see \fig~\ref{fig:vdm_design}) is a multi-NIC dual-mode design with five Intel AX200 NICs. Each NIC is configured via a helper framework~\cite{schepersFrameworkTestFuzz2021} to run both an AP interface and a monitor interface concurrently\footnote{For AX200, we apply a driver patch to enable promiscuous reporting on the AP interface, bypassing packet filtering applied to both modes by default.}\footnote{Putting all NICs in monitor mode alone is infeasible, since most NICs cannot transmit hardware-timed ACKs in monitor mode~\cite{schepersFrameworkTestFuzz2021}, nor can user space generate ACKs within the SIFS deadline~\cite{ieee80211n}. Without timely ACKs, authentication requests are endlessly retransmitted until failure.}. The five AP interfaces share the same SSID but expose unique BSSIDs, forming a standard Extended Service Set (ESS). This ensures reliable AP-Client association for WPA2-PSK validation, while the parallel monitor interfaces capture every uplink MPDU across the antenna array.

\head{RSS Snapshot Aggregation}
To construct a packet-level RSS vector from parallel NICs, we implement a custom aggregation pipeline (see \fig\ref{fig:vdm_design}). Each MPDU is identified by a \emph{composite key} consisting of the Source Address and Sequence Number (SN)\footnote{In practice, the 12-bit (4096) SN is unique within a short time window.}. A temporal buffer holds RSS reports for each key until either all NICs respond or a timeout occurs. At that point, the system finalizes an RSS vector, inserting placeholders (minimum RSS) for missing entries. This strategy balances completeness with latency, enabling real-time tracking while remaining robust to low SNR and packet loss.

\subsection{3D Printing Luneburg Lens Front-End}
\label{ssec:impl_luneburg_lens}

Fabricating a Wi-Fi \fLL for drone-aided WiSAR must satisfy four requirements: (i) a quasi-continuous GRIN profile with precisely controllable effective permittivity; (ii) a multi-wavelength aperture (\eg, >12~cm for 5~GHz bands) to mitigate diffraction effects; (iii) a lightweight yet robust structure suitable for on-drone mounting; and (iv) a low-cost, accessible, and easily reproducible process.

However, prior fabrication methods prove unsuitable for drone-aided WiSAR due to critical trade-offs: \emph{Stacked or drilled media}~\cite{borFoamBasedLuneburg2014,maThreedimensionalBroadbandBroadangle2010} create a stepped GRIN approximation and their multi-stage assembly introduces tolerance stack-up, leading to wavefront distortion. \emph{High-resolution stereolithography}~\cite{wuContinuousVariableDielectric2022} offers finely detailed structures, but suffers from higher costs and complex post-processing. \emph{Discrete FDM infills} like crossing structures~\cite{qianUniScatterMetamaterialBackscatter2023} result in coarse gradients, poor spherical symmetry, and limited mechanical strength.

We therefore adopt a monolithic lens fabricated via fused deposition modeling (FDM) 3D printing (see \fig\ref{fig:ll_inner}c), using \PLA\footnote{PLA is chosen for its permittivity stability across the 2.4/5~GHz Wi-Fi bands~\cite{zechmeisterComplexRelativePermittivity2019}, ensuring stable beam pattern across Wi-Fi bands.} material with a graded Gyroid~\cite{Schoen1970InfinitePM} infill, a type of Triply Periodic Minimal Surface (TPMS). Treating "air + polymer" as a mixture, the local effective permittivity follows the material volume fraction~\cite{qianUniScatterMetamaterialBackscatter2023}
\begin{equation}
\label{equ:eff_dc}
\varepsilon_{r,\mathrm{eff}}(r)=\alpha(r)\,\varepsilon_m+\big(1-\alpha(r)\big)\,\varepsilon_{r,\mathrm{air}},
\end{equation}
where $\varepsilon_m$ is the permittivity of PLA~\cite{zechmeisterComplexRelativePermittivity2019}, $\varepsilon_{r,\mathrm{air}}=1$, and $\alpha(r)$ is the local material volume fraction controlled by the Gyroid's parametric definition~\cite{al-ketanMultifunctionalMechanicalMetamaterials2019}, which results in a smooth radial sweep to realize the target GRIN (see \fig\ref{fig:ll_inner}a-b). This avoids stepped discretization while keeping the structure lightweight and stiff. 

We create the 3D model of the \fLL using \textit{MSLattice}~\cite{al-ketanMSLatticeFreeSoftware2021}. The unit-cell size is 1~cm (well below the Wi-Fi wavelength) with a lens radius of 7.5 cm, \revap{which is optimized for 5\,GHz Wi-Fi. While 2.4\,GHz benefits from better propagation, achieving equivalent gain requires a larger aperture\footnote{\revap{Despite the suboptimal size, experiments in \S\ref{sssec:exp_vdm} confirm the lens for 5\,GHz also extends the working range at 2.4\,GHz.}}, compromising drone mobility and durability.} To ensure manufacturability and structural integrity at the periphery, we truncated the target permittivity profile at 1.25 instead of the ideal 1.0, a necessary compromise to avoid near-zero material volume. We printed two variants on a consumer-grade FDM 3D printer~\cite{bambulabX1Carbon}: (i) the base Gyroid structure (\fig\ref{fig:ll_inner}c), and (ii) the same core with a uniform 0.5~mm outer skin to provide a solid surface for reliable \FPC antenna adhesion (\fig\ref{fig:impl} and \fig\ref{fig:measure_beam_pattern}c). This single-material process is highly cost-effective, with raw material costs of only 4 US dollars per 15~cm lens.

\begin{figure}[t]
    \centering
    \includegraphics[width=\linewidth]{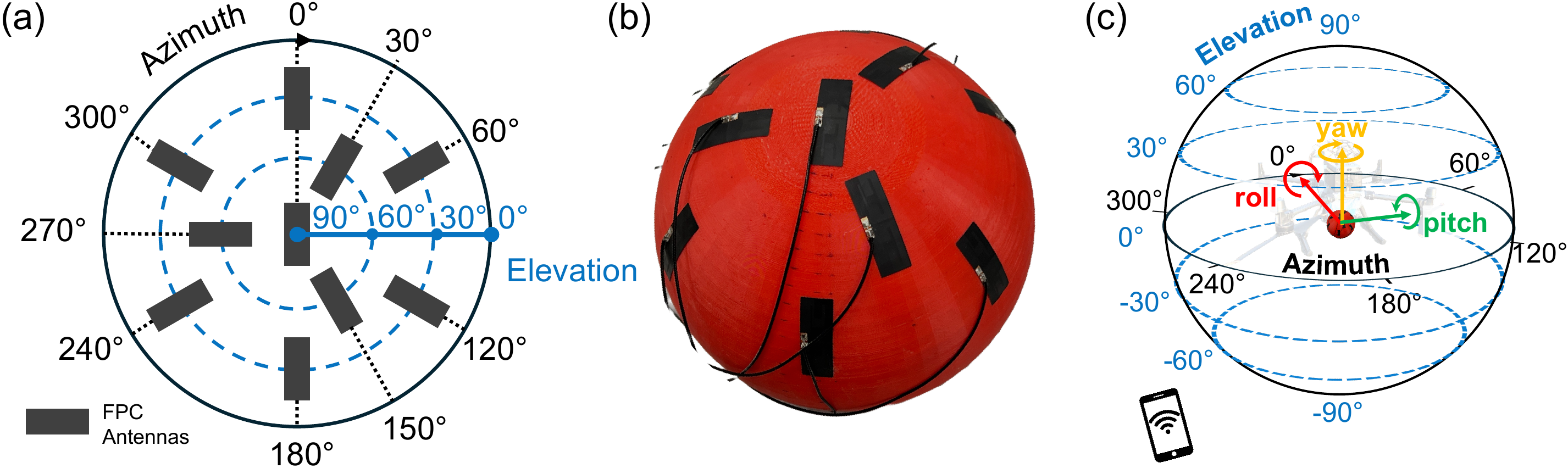}
    \caption{
    \rm{(a) FPC antenna layout. (b) Actual deployment. (c) Alignment with drone's coord. (roll, pitch, yaw).}
    }
    \label{fig:layout}
\end{figure}

\head{Beam Pattern Characterization and Re-usability}
To obtain the offline template $B_0$ for our AoA algorithm, we characterize the beam pattern of the fabricated \fLL assembly in a microwave anechoic chamber (see \fig\ref{fig:measure_beam_pattern}c). The measurement captures the composite response of the lens together with its surface-mounted FPC antennas. During the characterization, we use a commercial Wi-Fi transmitter operating at 5.745~GHz, while the \fLL with its $1\times3$~cm FPC antenna array acts as the receiver. RSS values are collected on a $10^\circ$ grid over both azimuth and elevation, and the results are spline-interpolated to a $1^\circ$ resolution for use in the estimator. The resulting beam pattern, shown in \fig~\ref{fig:measure_beam_pattern}b, exhibits a peak gain of approximately 14~dBi. The half-power beamwidth is about $60^\circ$ in both azimuth and elevation. This measured beam is broader than the $48^\circ$ beamwidth predicted by full-wave COMSOL simulations of the bare lens because the radiation pattern of the wide-beam FPC elements combines with the focusing effect of the lens, producing a wider composite lobe.

A critical requirement for our system is that this one-time, offline-calibrated pattern remains a \emph{reusable} template in all field deployments. The primary challenge in achieving this is mitigating near-field scattering from the drone's airframe, which is not present during the chamber measurement. Our solution is twofold. First, we physically decouple the lens from the drone by installing a layer of microwave-absorbing foam between the \fLL and the airframe (see \fig\ref{fig:impl}). This represents a practical trade-off, incurring minimal weight while improving the \emph{fidelity} of the lab-measured pattern in the operational environment. Second, the re-usability is further ensured by our algorithm's intrinsic robustness; the shape-based, de-meaned correlation estimator is inherently resilient to minor, real-world pattern variations. This combination of physical mitigation and algorithmic robustness allows the template to be reliably applied across missions without per-deployment recalibration.

\head{Antenna Layout}
We implement the antenna array on a 15\,cm Luneburg Lens for Wi-Fi bands, guided by two principles: maximizing information capture in the upper hemisphere, where the lens concentrates incident energy, and ensuring balanced angular sampling across both azimuth and elevation. To this end, our prototype employs ten antennas: one at the zenith for overhead sensitivity, and two concentric rings at $60^\circ$ and $30^\circ$ elevation, containing three and six elements, respectively. The $1{:}2$ allocation reflects the rings' circumferences, yielding near-uniform angular density and exploiting regions of highest beam gradient. As shown in \fig\ref{fig:layout}, this layout balances broad angular coverage with fine spatial resolution.

\subsection{Drone Integration}
A key aspect of the prototype implementation is integrating the \sysname payload with the drone's flight control system. This integration addresses the critical step of converting the estimated AoA into navigation commands. The raw AoA is estimated in the drone's local coordinate system (body frame), but for navigation, an absolute direction in the global coordinate system (world frame) is required. This conversion must account for the drone's constantly changing attitude.
To achieve this, we utilize the DJI Payload SDK~\cite{dji_payload_sdk} to continuously fetch the drone's real-time attitude from its onboard IMU. This data enables a rotational transformation to be applied to each raw AoA estimate, converting it from the relative body frame into a stable, world-referenced direction. The resulting compensated direction is then transmitted to the drone's flight controller via the existing communication link, enabling real-time guidance.

\section{Experiments}
\label{sec:exp}

We begin by outlining the experimental setup, followed by the definitions of key performance metrics. We then report a series of microbenchmarks and integrated system evaluations, concluding with a single-blind WiSAR trial where the location of the \emph{victim} remains unknown to the pilot. Our experiments and trials are approved by our institution's IRB and comply with local regulations. 
\begin{figure*}[t]
    \centering
    \includegraphics[width=\linewidth]{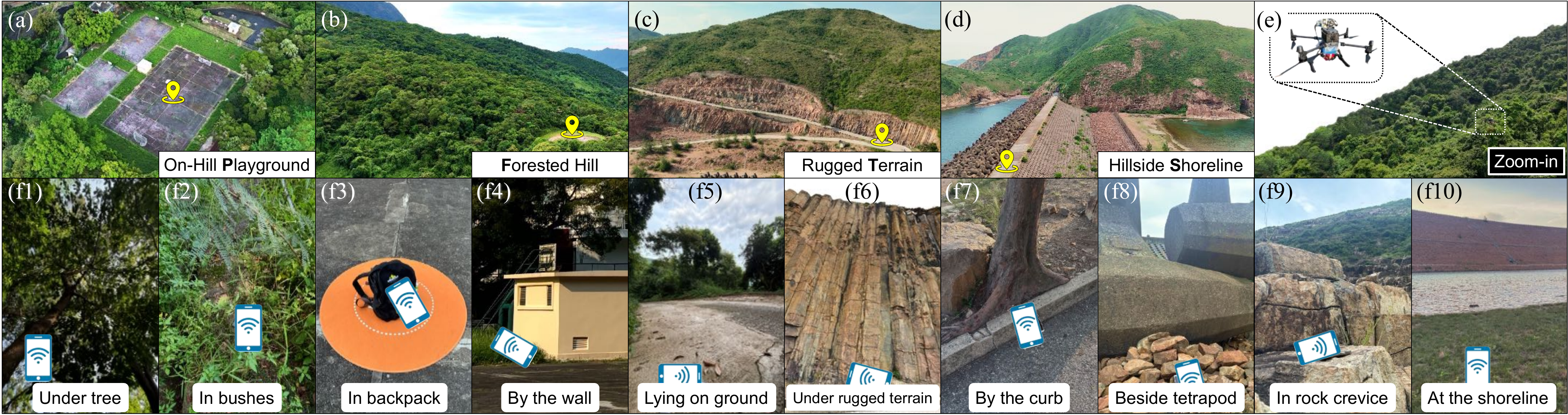}
    \caption{Experiment scenarios. {\rm(a)–(d) Bird's-eye views of four test environments: an on-hill playground, a forested hill, rugged terrain, and a hillside shoreline. (e) Zoom-in of the drone in operation. (f1)–(f10) typical placements for target devices.}
    }
    \label{fig:scenarios}
\end{figure*}

\subsection{Experimental Setup}
\label{ssec:exp_setup}

\head{Environments and Targets}
We evaluate \sysname across four distinct outdoor environments representing different levels of foliage coverage and terrain complexity (see \fig\ref{fig:scenarios}a-d): an on-hill \textbf{P}layground (Scene-P), a \textbf{F}orested hill (Scene-F), rugged \textbf{T}errain with exposed rock faces (Scene-T), and a hillside \textbf{S}horeline (Scene-S). Our targets are commodity Wi-Fi devices, including various smartphones, a tablet (iPad), and a smartwatch (Apple Watch), placed in realistic scenarios (see \fig\ref{fig:scenarios}f). The \emph{ground-truth} position for each target is obtained from its GPS coordinates, and refined against high-resolution satellite imagery to correct for GPS drift.

\head{Hardware}
Our primary evaluation platform is the \sysname prototype detailed in \S\ref{sec:impl}, which consists of a 15~cm \fLL and a ten-antenna array mounted on a DJI Matrice 350 RTK drone. The drone is either programmed to follow a pre-defined zigzag trajectory for systematic evaluation or manually piloted for realistic trials, following the instructions generated by drone navigation module and sent to the remote controller (see \fig\ref{fig:trajectory}). Unless otherwise specified, experiments are conducted on the 5\,GHz Wi-Fi band, with some tests including 2.4\,GHz for comparison.

\head{Baseline}
We benchmark \sysname's Direction Finding Module against \emph{ArrayTrack}~\cite{xiongArrayTrackFineGrainedIndoor2013}, a representative 2D AoA system based on Wi-Fi CSI. We implement it on a six-element Phaser-style ULA~\cite{gjengsetPhaserEnablingPhased2014} using five AX200 NICs on a Raspberry Pi with a modified driver for CSI extraction. For fairness, we also downgrade \sysname to a 2D six-antenna variant.

\head{Performance Metrics}
We use the following metrics to quantify the performance of individual components as well as the end-to-end system.

\noindent$\bullet$ \emph{Victim Discovery Range}: The maximum distance at which a target's packets can be detected and verified.

\noindent$\bullet$ \emph{Direction Finding Accuracy}: We quantify AoA accuracy using the \emph{Projection Rate} (PR), defined as $\text{PR} = \hat{\bs{u}} \cdot \bs{u}$, where $\hat{\bs{u}}$ is the estimated direction and $\bs{u}$ the ground truth. PR $\in [-1,1]$ reflects the effectiveness of motion along the estimated direction in reducing range: 1 = perfect alignment, 0 = orthogonal (no progress), negative = moving away. Unlike angular error, PR is signed and task-oriented. We report the \emph{MedPR} (median PR) as an indicator of typical performance.

\noindent$\bullet$ \emph{Exploratory Success Rate}: Fraction of target devices successfully connected during the exploratory search phase.

\noindent$\bullet$ \emph{Localization Error}: Final horizontal distance between the reported and ground-truth target positions.

\subsection{Component Performance}
\label{ssec:exp_micro}

We first evaluate the performance of \sysname's three core components in a series of experiments. 

\begin{figure}[t]
  \centering
  \begin{minipage}[t]{\linewidth}
    \centering
    \includegraphics[width=\linewidth]{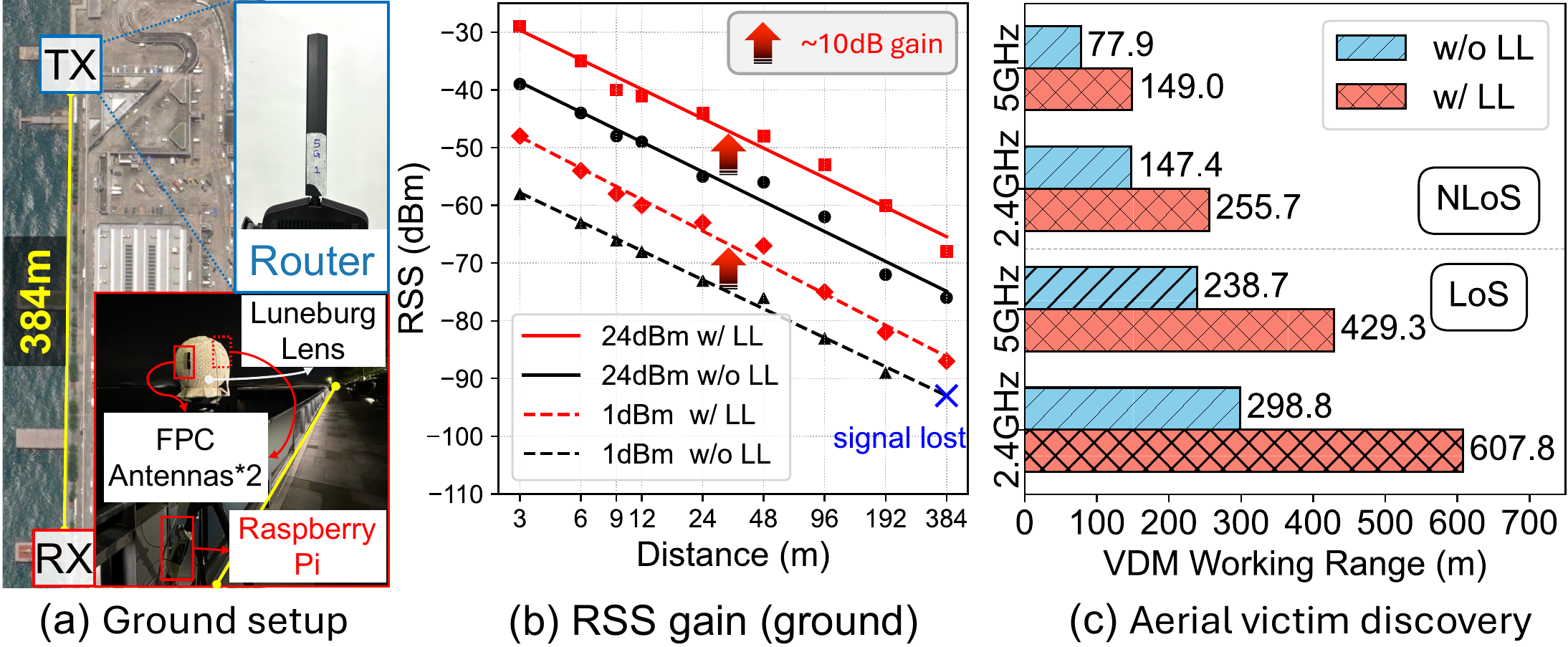}%
  \end{minipage}
  \caption{(a, b) RSS gain with Luneburg Lens (ground). (c) Extended victim discovery range (aerial).}
  \label{fig:ll_effectiveness}
\end{figure}

\subsubsection{Victim Discovery Module (VDM)}
\label{sssec:exp_vdm}
We first evaluate the \fLL's contribution to extending victim discovery range. In a ground test (see \fig\ref{fig:ll_effectiveness}a), a 5.745\,GHz router serves as the transmitter and a Raspberry~Pi with FPC antennas as the receiver. Placing the antenna at the lens focal point yields a RSS gain of about 10~dB across all tested distances for both high (24\,dBm) and low (1\,dBm) transmit powers, sufficient to raise weak signals above the noise floor and enable reception at 384\,m where the bare antenna fails (\fig\ref{fig:ll_effectiveness}b).
We then conduct aerial victim discovery trials in Scene-F to test whether this gain translates into operational range. As shown in \fig\ref{fig:ll_effectiveness}c, the \fLL extends detection range by 104\%/80\% in LoS at 2.4/5\,GHz, and by 73\%/91\% in NLoS, respectively. These results establish the lens as a key enabler that significantly broadens VDM's discovery range and boosts the likelihood of initial victim detection.

\begin{figure*}[t]
  \begin{minipage}{\linewidth}
     \subfloat[CDF of PR]{%
          \includegraphics[width=0.25\linewidth]{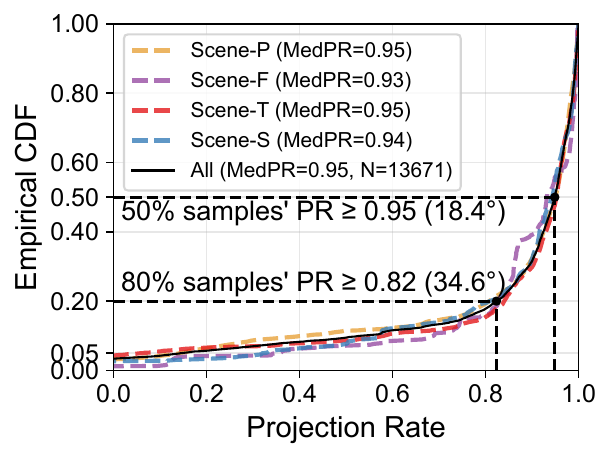}
            \label{fig:score_vs_scene}
    	  }
    \subfloat[Target placement]{%
          \includegraphics[width=0.25\linewidth]{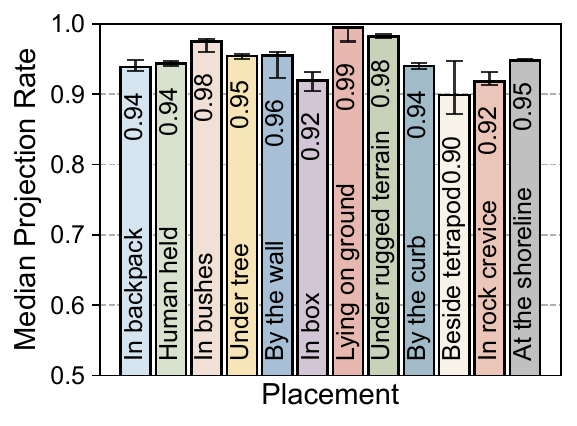}
        \label{fig:score_vs_placement}
            }
     \subfloat[Incident azimuth]{%
          \includegraphics[width=0.25\linewidth]{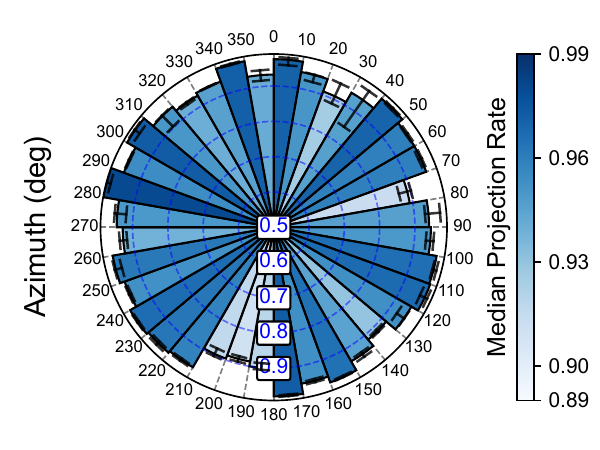}
            \label{fig:score_vs_azimuth}
    	  }
     \subfloat[Incident elevation]{%
          \includegraphics[width=0.25\linewidth]{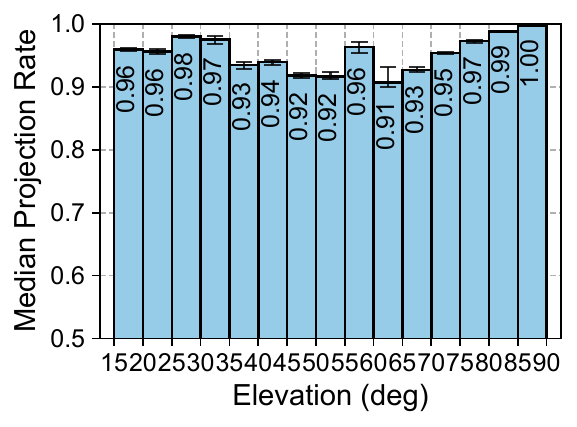}
            \label{fig:score_vs_elevation}
    	  }
    \newline
    \subfloat[Drone speed]{%
          \includegraphics[width=0.25\linewidth]{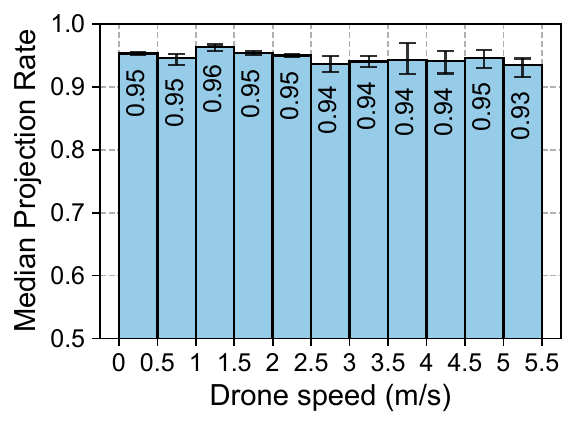}
        \label{fig:score_vs_speed}
            }
    \subfloat[Drone-Target distance]{%
          \includegraphics[width=0.25\linewidth]{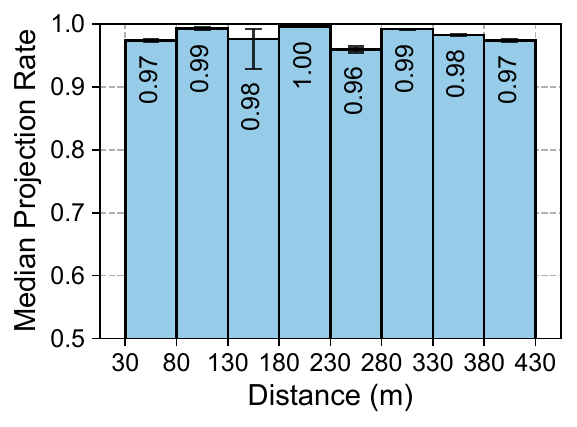}
            \label{fig:score_vs_distance}
    	  }
    \subfloat[Average RSS]{%
          \includegraphics[width=0.25\linewidth]{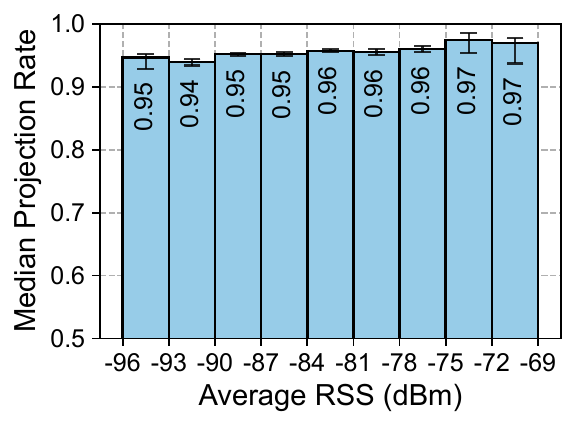}
            \label{fig:score_vs_avg_rss}
    	  }
    \subfloat[LL vs. ArrayTrack]{%
          \includegraphics[width=0.25\linewidth]{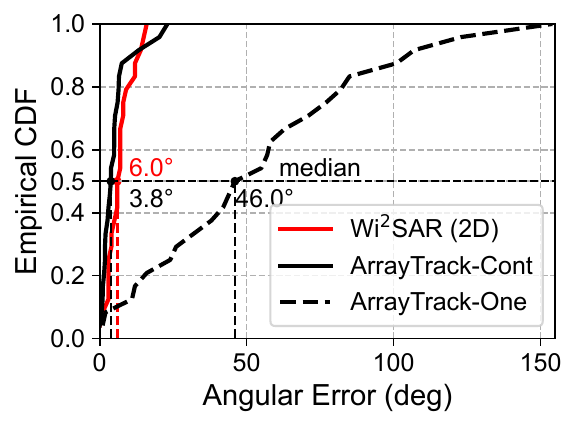}
            \label{fig:ll_vs_arraytrack_cdf}
    	  }
    \caption{
    Performance of Direction Finding Module. \rm{\textbf{(a)} Overall performance of four scenes. \textbf{(b)-(g)} Impact of diverse conditions. Error bars indicate 95\% bootstrap confidence intervals. \textbf{(h)} Comparison study with ArrayTrack.
    }}
    \label{fig:performance_df}
    \end{minipage}
\end{figure*}

\subsubsection{Direction Finding Module (DFM)}
\label{ssec:exp_df}

We first quantify the overall accuracy of the DFM across diverse scenarios, then examine its robustness to impact factors, and finally benchmark it against the CSI-based baseline.

\begin{figure*}[t]
    \centering
    \includegraphics[width=\linewidth]{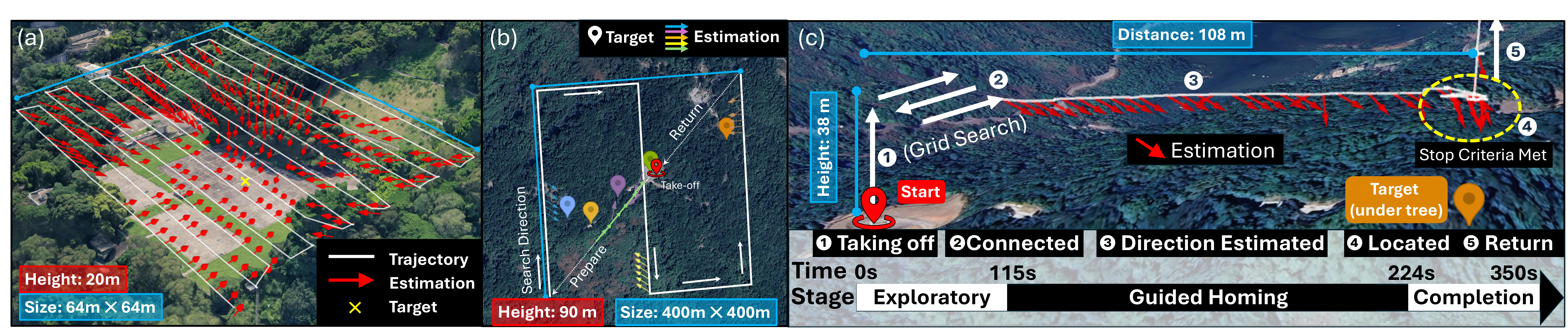}
    \caption{
    Example Drone Trajectories across Trials. \rm {(a) Controlled zigzag flight with angle predictions. (b) Large-area exploratory search (400m × 400m). (c) Full WiSAR trial showing grid search, followed by direction-guided search, localizing the target under foliage within 224 s.
    }
    }
    \label{fig:trajectory}
\end{figure*}

\head{Direction Finding Performance}
We conducted controlled drone flights across four environments with varying vegetation density and terrain evenness. Each received packet is treated as an independent sample, yielding 13,671 samples in total from heterogeneous targets including smartphones, tablets, and wearables, under 12 common placements and drone speeds ranging from hovering to 5.5\,m/s (example trajectory in \fig\ref{fig:trajectory}a). As shown in \fig\ref{fig:score_vs_scene}, \sysname consistently delivers reliable 3D direction finding: the MedPR reaches 0.95, corresponding to an 18.4$^\circ$ angular error, while the 80\%-tile angular error is 34.6$^\circ$. Fewer than 4\% of the samples fall into ambiguous cases (orthogonal or opposite bearings). These results confirm that DFM achieves high accuracy and stable performance across diverse terrains, device types, user placements, and drone maneuvers.

\noindent $\bullet$ \textit{Target Placement.} Victims may carry devices in diverse positions during a rescue scenario. We therefore test 12 typical placements in WiSAR. As shown in \fig\ref{fig:score_vs_placement}, even under near-device occlusion that interferes with the signal, the DFM maintains MedPR above 0.9, which is sufficient to reliably guide the drone.

\noindent $\bullet$ \textit{Incident Angle.} Since signals may arrive from arbitrary directions, we examine incident angles to assess directional consistency. Across azimuth directions, MedPR remains high between 0.89 and 0.99 (\fig\ref{fig:score_vs_azimuth}). Along elevation (see \fig\ref{fig:score_vs_elevation}), performance exhibits clear peaks around 30$^\circ$, 60$^\circ$, and zenith, which align with the antenna rings in our implementation (\S\ref{ssec:impl_luneburg_lens}). Moreover, accuracy progressively improves as elevation approaches 90$^\circ$, ensuring that our stop criterion converges reliably during search.

\noindent $\bullet$ \textit{Drone Speed.} Drone motion can introduce vibration and channel dynamics that challenge direction finding. Testing speeds from hovering to over 5\,m/s, we observe stable MedPR between 0.93 and 0.96 (\fig\ref{fig:score_vs_speed}), indicating robustness to motion-induced fluctuations across both slow maneuvers and fast scans.

\noindent $\bullet$ \textit{Distance and Attenuation.} 
Search operations often require long-range detection under weak signals. \sysname sustains high accuracy with MedPR $\ge 0.96$ at distances up to 430\,m (\fig\ref{fig:score_vs_distance}). More importantly, performance degrades gracefully with signal attenuation: even when mean RSS drops to $-93\,\text{dBm}$, the DFM still achieves a MedPR of 0.94 (\fig\ref{fig:score_vs_avg_rss}), demonstrating robustness under low-SNR conditions.

\noindent $\bullet$ \textit{Device Types.} 
Rescue targets may use heterogeneous devices with different transmit powers. Without any algorithm modification, our \sysname generalizes well across various devices, achieving high MedPR and long detection ranges in LoS conditions: smartphones (429\,m, MedPR 0.951), a tablet (152\,m, MedPR 0.918), and even a low-power smartwatch (197\,m, MedPR 0.939).

\head{Comparison with Baselines}
CSI-based methods such as \texttt{ArrayTrack} require continuous phase calibration to correct hardware-induced, time-varying phase offsets~\cite{gjengsetPhaserEnablingPhased2014,zhuCalibratingTimevariantDevicespecific2017}. While effective in controlled ground experiments, such calibration is impractical for drones in WiSAR scenarios, where a static reference signal is difficult to obtain in flight.
To highlight this limitation, we evaluate two ArrayTrack variants: \emph{ArrayTrack-Cont}, which continuously recalibrates using a static reference signal, and \emph{ArrayTrack-One}, which performs only a one-time calibration prior to each deployment without online updates. For fairness, we use a 2D variant of \sysname restricted to azimuth, aligning with ArrayTrack's 2D setup\footnote{Both \emph{ArrayTrack} and \sysname use six antennas in our experiments. ArrayTrack's field of view (FoV) is inherently limited to $180^\circ$ due to its linear array design; thus we report its accuracy only within this span, while \sysname provides and is evaluated over the full $360^\circ$ FoV.}. As shown in \fig\ref{fig:ll_vs_arraytrack_cdf}, \emph{ArrayTrack-Cont} achieves $3.8^\circ$ median error but requires an impractical reference signal, while \emph{ArrayTrack-One} degrades to $46.0^\circ$ once the pre-deployment
calibration becomes invalid. In contrast, \sysname requires only a single design-time characterization of its amplitude profile; this profile remains stable across hardware instances and deployments, sustaining $6.0^\circ$ median error across the full $360^\circ$ FoV without any recalibration, highlighting both its robustness and deployability for drone-based WiSAR.

\subsubsection{Search Scheme Performance}
\label{ssec:exp_navigation}
\revap{We evaluate \sysname's performance in a mid-scale feasibility study of exploratory search. The test area measures 400\,m $\times$ 400\,m (160,000 m$^2$)\footnote{\revap{The area size was constrained by local flight regulations that mandate line-of-sight operations in forested zones.}}. In this area,} five phones are placed at unknown locations, and the drone executes a zigzag search at 2 m/s. As shown in \fig\ref{fig:trajectory}b, \sysname successfully discovered and provided initial direction estimates for all five targets within \rev{13.5} minutes, achieving a 100\% discovery rate in this large environment. By contrast, covering a comparable area typically requires multiple ground searchers working for several hours.
\revap{%
We further analyze the discovery latency, defined as the time elapsed from when the drone enters the victim's potential communication range (see \fig\ref{fig:ll_effectiveness}c) until the first successful connection. Based on log data from the five test phones, the average discovery latency is 39.8\,s. This latency is primarily determined by the victim device's Wi-Fi scan interval. Our offline measurements show that the median scan intervals of these devices in active, idle, and power-saving modes are 9\,s, 36\,s, and 165\,s, respectively. Consequently, flying faster is not always better: if the drone passes through the working range too quickly, the victim may not scan and auto-connect during that visit. To mitigate this, one can (i) reduce the flight speed, (ii) use a denser zigzag grid to increase overlap and revisit frequency, or (iii) deploy multiple drones to shorten revisit times, improving the chance of timely discovery.}

\subsection{End-to-End Trial: \sysname in the Wild}
\label{ssec:exp_casestudy}
We conduct a single-blind WiSAR trial in the foliage-dense Scene-F, where the pilot has no prior knowledge of the target and relies solely on \sysname's guidance, searching a forested area of 40,000 m$^2$. The full trajectory is shown in \fig\ref{fig:trajectory}c. The drone first performs an exploratory search and acquires the initial signal, \ie, the target's auto-reconnection request, at 115 s, then switches to guided navigation. At 224 s, the system signals successful localization; the pilot reports the target's position by ascending, and we log the drone's GPS coordinates at that moment. The resulting horizontal localization error is only \textbf{5 m}. This trial demonstrates that \sysname efficiently and accurately guides search operations to completion under realistic conditions.

\subsection{System Latency and Overhead}
\label{ssec:sys_latency}
We profile \sysname's runtime performance on the Raspberry Pi CM4 used onboard our drone: Processing a single RSS snapshot for direction finding takes 48 ms, and, with seven concurrent targets the system sustains an overall update rate of 7.8 Hz, sufficient for real-time tracking. The pipeline consumes 635 MB memory and 48–52\% CPU, confirming that \sysname runs efficiently on embedded hardware.
\revap{%
Additionally, we quantify the impact of the \sysname payload on drone battery endurance. During the 13.5-minute search flight in \S\ref{ssec:exp_navigation}, the drone consumed an extra 6\% battery with the payload compared with a matched flight without the payload at the same speed and duration, indicating a manageable impact for practical WiSAR missions.}

\section{Discussions and Future Work}
\label{sec:discuss}
As a pioneering Wi-Fi drone system for WiSAR, there is room to further improve \sysname.

\head{Random Targets}
In cases where the target is an unidentified individual (\eg, a missing hiker with no prior information), \sysname can be adapted to broadcast common network beacons (\eg, airport Wi-Fi) and overhear all nearby network traffic. In a wilderness environment where such traffic is sparse, any active device becomes a potential point of interest for investigating all signs of digital life.

\head{Scalability to Larger Areas}
\revap{While our experiments validate \sysname in a mid-scale area (400\,m $\times$ 400\,m), the design is scalable to larger regions. By partitioning a large search area into manageable grids, our proven mid-scale search performance can be replicated across each grid section. This "divide and conquer" approach can be further accelerated by employing multi-drone swarms to search multiple grids in parallel, sharing discoveries for expedited search. We are currently engaging with a local voluntary mountain search team for larger-scale field trials in real-world search.}

\revap{\head{Privacy and Ethics} \sysname should be deployed only by authorized organizations with appropriate consent or authorization. Unlike LTE pseudo-base stations~\cite{albaneseSARDOAutomatedSearchandRescue2022} that can use persistent identities (IMSI), \sysname uses changeable Wi-Fi credentials (SSID/PSK) for the mission; discard all credentials and collected traffic after the mission. Specific legal and ethical adjudication regarding this controlled application is left to regulatory bodies.}

\head{Implications for Wi-Fi protocol} \sysname's reliance on auto-reconnection behaviors highlights potential modifications to Wi-Fi authentication protocols that could improve both security and usability in emergency scenarios. Future iterations of IEEE 802.11 could consider standardized support for emergency beacon SSIDs, enabling sanctioned rescue operations to trigger safe and rapid device response.

\head{\revap{Migration Beyond Wi-Fi Signals}}
The core principle of \sysname could be extended beyond Wi-Fi signals. The Luneburg Lens, being a passive electromagnetic lens, could be adapted for other \revap{wireless signals like Cellular (LTE/5G/6G), LoRa, or Bluetooth}. 
\revap{The migration requires three key steps: (1) adopting a "spoofing" technique for packet elicitation and identification; (2) adapting the Luneburg Lens to the new wavelength; and (3) re-characterizing the beam pattern. Since \sysname relies only on RSS rather than specific protocols, our open-sourced direction-finding algorithm can be directly reused to process the captured RSS patterns.}

\head{Victims without Wi-Fi}
There are cases where a victim does not have a powered Wi-Fi device and \sysname is not applicable. However, we believe a system that can facilitate rescue and save lives in a significant subset of cases is already invaluable. Furthermore, the system could be extended to work with dedicated low-power electronic trackers for hikers, reinforcing its effectiveness.

\section{Related Works}
\label{sec:related_works}
\noindent\textbf{Drone-aided WiSAR.}
In the domain of WiSAR operations, drones have emerged as potent tools to augment manual search efforts~\cite{lyuUnmannedAerialVehicles2023}. Various systems employ vision-based sensors such as RGB cameras~\cite{andriluka2010vision,tusnioEfficiencyDronesUsage2021}, infrared and thermal cameras~\cite{leira2021object, de2018using,chretien2015wildlife}. 
\revap{However, vision-based pipelines degrade severely under canopy or rugged rocky areas, despite efforts to address partial occlusion~\cite{schedlAutonomousDroneSearch2021, broylesWiSARDLabeledVisual2022}. In practice, RGB and thermal imaging are well-suited for rapid sweeps over open terrain, but their effective working range drops sharply once occlusion becomes dominant.}
To mitigate weather and occlusion, RF sensing has been explored: recent work attaches mmWave radar to drones to detect subtle chest motion as a life sign~\cite{zhangRFSearchSearchingUnconscious2023}, \revap{demonstrating promise \emph{indoors}, but \emph{outdoor} dynamics easily mask such weak physiological signals at long ranges.}
Beyond vision and mmWave, several attempts localize mobile devices via infrastructure-free airborne radios, such as LTE pseudo-base-station methods~\cite{albaneseSARDOAutomatedSearchandRescue2022} and non-cooperative Wi-Fi ranging~\cite{hoOpenCollaborativePlatform2022,abediNoncooperativeWifiLocalization2022a,wangFeasibilityStudyMobile2013,acunaLocalizationWiFiProbe2017,sunLocalizationWiFiUAV2018}. 
\revap{However, most of these approaches rely on range-based estimators that require multiple measurements from different positions. Without directional guidance, the drone may inadvertently fly away from rather than toward the target, wasting both time and energy or even failing the search task.}

\noindent\textbf{\revap{Wi-Fi localization for WiSAR.}}
\revap{Among RF modalities, Wi-Fi stands out for its ubiquity; however, adapting indoor positioning techniques to wilderness settings is challenging.} Classic Wi-Fi localization spans three main categories.
First, fingerprinting~\cite{yangLocatingFingerprintSpace2012} builds a radio map from dense site surveys, which is impractical in wilderness where pre-survey is impossible and propagation varies widely.
Second, distance-based methods estimate range via RSS path-loss~\cite{bahlRADARBuildingRFbased2000,youssefHorusWLANLocation2005} or ToF~\cite{vasisht2016decimeter}. RSS trilateration is brittle in WiSAR due to foliage absorption and multipath, while ToF approaches require client cooperation or nanosecond-scale synchronization.
Third, direction-based methods estimate AoA from phase differences with subspace estimators~\cite{xiongArrayTrackFineGrainedIndoor2013,kotaruSpotFiDecimeterLevel2015,gjengsetPhaserEnablingPhased2014}. These techniques achieve high accuracy indoors but depend on adequate SNR, tight inter-element phase coherence, and careful phase calibration. Such assumptions break down on vibrating aerial platforms operating near the noise floor, and few works~\cite{wangWiCALAccurateWiFiBased2025,Zhang20193DWiFi3L} demonstrate true long-range 3D bearing suitable for \revap{drone} navigation.
To the best of our knowledge, \sysname is the first system for RSS-only 3D direction finding on a Wi-Fi-enabled drone over extended distances, with non-cooperative Wi-Fi devices and free from on-the-fly calibration. Our approach differs from traditional RSS-based localization, utilizing the Luneburg Lens to boost signal \revap{strengths} and find direction from unique RSS patterns, and expands the applications to outdoor WiSAR.

\section{Conclusion}
\label{sec:conclusion}
This paper presents the design and implementation of \sysname, a novel drone-based wireless system for automatic WiSAR operations without relying on any existing infrastructure. 
\sysname leverages the auto-reconnection behavior of Wi-Fi networks to discover and locate victims by their accompanying devices. 
\sysname incorporates three core components: a power-efficient victim discovery scheme, a long-range 3D direction finding approach using a 3D-printed Luneburg Lens, and a direction-guided drone navigation strategy. 
We implement \sysname as an end-to-end system on commodity drones and validate it in real-world wilderness scenarios. 
The results highlight its remarkable performance, underpinning its potential for real-world deployment.

\begin{acks}
    This work was supported by NSFC under grant No. 62222216, Hong Kong RGC GRF No. 17212224 and No. 17211725, and CRF No. C5002-23Y. We thank the anonymous Reviewers and our Shepherd for their insightful feedback. 
\end{acks}

\bibliographystyle{ACM-Reference-Format}
\bibliography{zotero-reference, manualAdd-refs}

\end{document}